\documentclass[12pt, draftclsnofoot, onecolumn]{IEEEtran}
\usepackage{graphicx}
\usepackage{caption}
\ifCLASSOPTIONcompsoc
  \usepackage[caption=false,font=normalsize,labelfont=sf,textfont=sf]{subfig}
\else
  \usepackage[caption=false,font=footnotesize]{subfig}
\fi
\usepackage{indentfirst}

\usepackage{amsmath}
\usepackage{amssymb}
\usepackage{bbm}
\usepackage{times}
\usepackage{mathtools}
\usepackage{psfrag}
\usepackage{cite}
\usepackage{lastpage}
\usepackage{fancyhdr}
\usepackage{color}
 \usepackage{amsthm}
\usepackage{bigints}
\newtheorem{theorem}{Theorem}
\newtheorem{Proposition}{Proposition}
\newtheorem{Lemma}{Lemma}
\newtheorem{Remark}{Remark}
\newtheorem{Corollary}{Corollary}
\begin{document}
\title{A {\color{black}User-Centric Cooperative} Scheme for UAV Assisted {\color{black}Wireless} Networks in Malfunction  {\color{black}Areas}}
\author{Yanshi Sun, Zhiguo Ding, \IEEEmembership{Senior Member, IEEE}, Xuchu Dai,
\thanks{Y. Sun and X. Dai are with the Key Laboratory of Wireless-Optical Communications, Chinese Academy of Sciences, School of Information Science and Technology, University of Science and Technology of China,
No. 96 Jinzhai Road, Hefei, Anhui Province, 230026, P. R. China. (email: sys@mail.ustc.edu.cn, daixc@ustc.edu.cn).

Z. Ding is with the School of Electrical and Electronic Engineering, the University of Manchester, Manchester M13 9PL, U.K. (email:zhiguo.ding@manchester.ac.uk).
}}
\maketitle
\begin{abstract}
A promising application of unmanned aerial {\color{black}vehicles (UAVs)} to the future communication networks
is to address emergency communications.
This paper considers such a scenario where a UAV is employed to a malfunction area
(modeled as a {\color{black}circular disc}) in which all ground base stations (BSs) break down. The ground BSs outside the malfunction area are modelled as {\color{black}a} homogeneous
Poisson point process (HPPP).
Particularly, {\color{black}a user-centric cooperative scheme is proposed to serve the UEs in the malfunction area}.
{\color{black}According to the user equipment's (UE's) connections to the UAV and the nearest ground BS, the malfunction
area is divided into three regions, namely the UAV region, the cooperation region and the nearest ground BS region, in which the UEs are served by the UAV only, both the UAV and the nearest ground BS, and the nearest ground BS, respectively.}
The region size of each type can be adjusted by a cooperation parameter $\delta$.
Through rigorous derivations, {\color{black}an} expression for the coverage probability achieved by
the UE in the malfunction area is obtained.
In order to provide {\color{black}a} fair comparison, {\color{black}the} normalized spectral efficiency (NSE) {\color{black}which is} defined by taking both system throughput and the number of serving BSs into consideration, {\color{black}is used as a criterion for the performance evaluation}.
Numerical results are presented to verify the accuracy of the analytical results and also
to demonstrate the superior performance of the proposed scheme.
\end{abstract}
\begin{IEEEkeywords}
Stochastic geometry (SG), unmanned aerial vehicle (UAV), cooperative communication,
coverage probability, emergence communication.
\end{IEEEkeywords}
\section{Introduction}
\IEEEPARstart{R}ecently,  with {\color{black}a significant} improvement of drone {\color{black}technologies},
such as increased payload capacity, prolonged flight endurance{\color{black}, etc}, the
application of unmanned aerial vehicles (UAVs) has attracted extensive attentions
in both academia and industry \cite{valavanis2015future,asadpour2014micro}.
One promising application is the use of UAV as flying base stations (BSs), which aims to
boost the capabilities of the existing terrestrial cellular networks \cite{zeng2016wireless,xiao2016enabling,menouar2017uav,chen2018multiple}.
One key feature of UAVs is their agility and mobility. For example, UAVs can be deployed
in a very short time with {\color{black}a} fairly low cost compared to the deployment of
traditional terrestrial BSs.
Moreover, UAVs have the ability to intelligently adjust their positions in real-time to
efficiently provide {\color{black}large coverage and improve quality of} certain links.
Another feature which makes UAVs appealing is the higher opportunity to provide line-of-sight
(LoS) links, which can potentially provide more reliable links for certain users and hence
provide better quality of service (QoS), {\color{black}compared to traditional terrestrial BSs} .
Due to the above advantages, UAVs can be applied to various particular scenarios
for future communication networks.
One application scenario is to address temporary events such as concerts and sporting events,
where excessive connectivity and rate requirements are demanded by a large number of
audience.
Besides, in some unexpected scenarios such as disasters and emergency accidents,
terrestrial networks may be {\color{black}broken down} due to equipment damage or power failure, UAVs can
play an important role to help to reconstruct communication quickly and efficiently \cite{merwaday2016improved,kandeepan2014aerial,chandrasekharan2016designing}.
Other potential application scenarios include Internet of Things (IoT) \cite{mozaffari2017mobile}, public safety networks \cite{merwaday2015uav}, mobile edge computing \cite{jeong2018mobile} etc.
\subsection{Related works and Motivation}
To realize the application and reap the benefit of UAVs {\color{black}in} future communication networks,
researchers have done great efforts to address various technical challenges
including but not limited to channel modeling, deployment problems, trajectory design,
resource management and  performance analysis, {\color{black}as illustrated in the following}.

Air-to-ground channel modeling is an important part of the existing work on UAV technologies.
In \cite{matolak2015unmanned}, simulation and measurement results for path loss, delay spread
and fading in air-to-ground channels were presented.
It has been shown in \cite{feng2006path} and \cite{holis2008elevation} that the characteristics
of the air-to-ground channel are dependent on the height of the aerial BSs, because of path loss and
shadowing.
In \cite{al2014conf} and \cite{al2014letter}, the authors studied the impact of
environment parameters on air-ground channel path loss and then proposed {\color{black}an} elevation angle
dependent function to characterize the probabilities of LoS and NLoS links between a low
altitude platform and a ground device.
The second {\color{black}important} research direction of UAV is to solve optimization problems which {\color{black}are}
relevant to UAV parameters, such as deployment \cite{mozaffari2017mobile}, cellular network planing with UAVs \cite{sharma2016uav},
trajectory optimization \cite{xu2018uav,pang2018uav} and resource management \cite{lyu2016cyclical}.

Another important research direction, which is complementary to the above two kinds of work,
is to {\color{black}carry out the} system-level performance {\color{black}evaluation} by utilizing tools from stochastic geometry \cite{haenggi2012stochastic}.
This kind of work  usually aims to evaluate the impact of main design parameters on the
system performance and reveal the hidden tradeoffs when designing UAV assisted networks
\cite{mozaffari2016unmanned,zhang2017spectrum,ye2018secure,chetlur2017downlink,turgut2018downlink
,wang2018modeling,wang2018modeling2}. For example, the authors in \cite{mozaffari2016unmanned} studied
the downlink coverage and rate performance of a single UAV that co-exists with a device-to-device
(D2D) communication network. The authors in \cite{zhang2017spectrum} used 3D Poisson point process (PPP) to
analyze the performance of a network composed by UAVs and underlaid conventional cellular networks.
In \cite{chetlur2017downlink}, the authors studied the performance achieved by ground users served
by multiple UAVs in a finite area, by using {\color{black}the binomial} Poisson process (BPP) model. Later, the
authors in \cite{wang2018modeling} extended the work in\cite{chetlur2017downlink} by taking
PPP modeled ground BSs into consideration.
In \cite{turgut2018downlink}, the authors provided {\color{black}an} analytical framework to analyze the
performance of UAV assisted cellular networks with clustered user equipments (UEs).

Different from the existing work in the literature for performance analysis, the authors
in \cite{wang2018modeling2} considered a scenario where a UAV hovers over the center of a
malfunction area (modeled as a {\color{black}circular disc}) to provide service to the UEs {\color{black}within the disc}.
Specifically, all ground BSs within the malfunction area break down, while {\color{black} those outside ground BSs}
work well {\color{black}and can be} modeled as points of a PPP removing the circular malfunction area.
{\color{black}It is important to point out that the work in \cite{wang2018modeling2} requires an assumption that} all UEs in the malfunction area are served by the UAV, which is not practical for UEs
locate {\color{black}in} the middle and edge areas of the malfunction area.
Intuitively, it is better to serve a UE {\color{black}in} the edge area by a ground BS outside the malfunction area
{\color{black}instead of the UAV in order to avoid strong path loss}. Besides, a UE locates {\color{black}in} the middle area is better to be
cooperatively served by the UAV and a ground BS, because the UE is relatively far from both
the UAV and ground BSs. The above observations reveal the importance of introducing cooperative
transmission schemes for the considered scenario, which motivates the work {\color{black}in} this paper.
\subsection{Contributions}
The main contributions of this paper are listed as follows.
\begin{itemize}
  \item By considering the same system model as used in \cite{wang2018modeling2}, this paper
  proposes a novel user-centric cooperative transmission scheme.
  In the proposed scheme, {\color{black}a} UE  chooses to be served by the UAV only, the nearest ground BS only,
  or both the UAV and the nearest ground BS, depending on the relationship between
  the average received power from the UAV and the nearest ground BS.
  Hence there are three kind of {\color{black}UEs}.
  The proportion of each kind of {\color{black}UEs} in the malfunction area can be tuned by a
  cooperation parameter $\delta$, ranging from zero to one.
  The significance of the proposed scheme is that it not only improves the coverage performance
  achieved by the UE compared to the scheme in \cite{wang2018modeling2}, but also {\color{black}takes the number of serving BSs into consideration}.
  \item It is necessary to point out that the proposed scheme in this paper is inspired
  by the work in \cite{feng2018tunable}, where {\color{black}a} tunable cooperation scheme {\color{black}was} proposed for a
  PPP based cellular network.
  However, {\color{black}the scenario considered in \cite{feng2018tunable} is different from the one
  in this paper}, which complicates the design of the transmission scheme.
  For example, in this paper, the probabilistic LoS/NLoS propagation model is used to characterize the air-to-ground
  channel which is different from traditional ground-to-ground channels.
  Since the propagation features of {\color{black}an} LoS link and {\color{black}an} NLoS link are different, the corresponding
  transmission strategies {\color{black}also become} different. Thus this paper uses the average received
  power as the measure, instead of the distance as used in \cite{feng2018tunable},
  to decide which transmission strategy should be used.
  \item Coverage probability achieved by a random UE in the malfunction area is used as one of the metrics to evaluate the performance of
  the proposed scheme.
  There are two main difficulties  to evaluate the
  coverage probability, which makes the analytical development challenging.
  The first is to
   derive the distribution for the distance from the UE to the nearest ground BS.
   The derivation here is not as easy as that in typical 2D PPP based models, due to
   the constraint that the ground BSs reside outside the malfunction area.
 {\color{black}The second is to obtain the Laplace transform of the aggregated interference from the ground BSs which are farther than the nearest ground BS, and the corresponding derivatives of the Laplace transform.}
 The difficulty here is that, the Laplace transform which  {\color{black}is}  to derive is dependent
 on the  relationship  {\color{black}between} the distance to the origin, the distance to the nearest ground BS
 and the radius of the circular malfunction area, which significantly complicates the geometric
 manipulation.
 Normalized spectral efficiency is also used as a metric to evaluate
 the performance, which takes both the system throughput and  {\color{black}the} number of serving  {\color{black}the} BSs
 into consideration.
 \item Analytical results are verified by computer simulations. To get insight into the proposed scheme, the impact of system parameters, such as UAV altitude, cooperation parameter $\delta$ and ground BS density etc,  is discussed. Two benchmark schemes are considered to facilitate comparison.
     One is the scheme used in \cite{wang2018modeling2}, where the UE in the malfunction area is served by the UAV only. The other is the  {\color{black}case where} there's no UAV deployed in the area and the UE is only served by the nearest ground BS outside the area. {\color{black}The provided comparison} results demonstrate the superiority of the proposed scheme over the above two benchmarks.
\end{itemize}

The rest of this paper is organized as follows. Section II illustrates the considered system model and
presents the transmission scheme.
Section III develops the analysis for the coverage probability achieved by a UE.
Section IV provides numerical results to demonstrate the performance of the proposed scheme
and also verify the accuracy of the developed analytical results.
Section V concludes the paper.
Finally,  {\color{black}appendixes} collect the proofs of the obtained analytical results.

\section{System Model}
\subsection{Location description}
Consider a downlink cellular network, where the ground BSs are randomly distributed in the plane.
Particularly, the locations of the ground BSs are modeled as a PPP, which is denoted by $\Phi$ with intensity $\lambda$.
There is an isolated region which is modeled as a disc $\mathcal{D}$ with radius $R_c$. Without loss of generality,
the center of the disc is set at the origin. It is assumed that,  {\color{black}because of  natural disaster or regional power failure}, all the ground BSs in disc  $\mathcal{D}$ break down and are disabled to serve.
The locations of the remaining ground base stations outside disc $\mathcal{D}$ are denoted by $y_i$, forming a new point process $\Phi_o$, where $\Phi_o=\Phi \backslash \mathcal{D}$.

\begin{figure}[!t]
\centering
\includegraphics[width=3.5in]{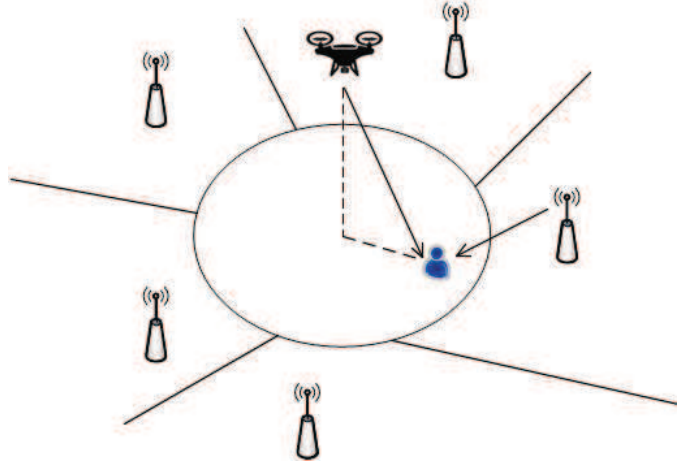}
\caption{An illustration of the system model.}
\label{system_model}
\end{figure}

As in \cite{wang2018modeling2}, a UAV is employed to address the emergency, which hovers at altitude $H$ at the center of disc $\mathcal{D}$.
This paper focuses on the performance of the UEs in the malfunction area $\mathcal{D}$.
Particularly, consider a UE, as shown in Fig. \ref{system_model},
the horizontal distance between the UAV to the UE is denoted by $r_0$.
Without loss of generality, the coordination of the UE is denoted by $x_0=(r_0,0)$.
The distance between the $i$-th ground base station to the UE is denoted by $r_i$, i.e.,
$r_i=||y_i-x_0||$.
Note that, the ground BSs are ordered according to their {\color{black}distances} to the UE,
i.e., $r_i\leq r_j$ ($0<i\leq j$).

\subsection{Channel model}
Note that, there are two kinds of channels in the considered scenario.
The first is the channel between the UAV and the UE, namely the air-to-ground channel.
The second is the channel between a ground BS and the UE,
namely the ground-to-ground channel.

To model the air-to-ground channel, the following two observations {\color{black}are worth being}
noticed.
On the one hand, note that an appealing feature of deploying UAV the increased
possibility of serving a UE through {\color{black}an} LoS link, which experiences lower propagation attenuation
than an NLoS link.
On the other hand, it is usually inevitable that the link between the UAV and the UE
is an NLoS link, due to the blockage effect caused by building, trees, etc.
To take the above two observations into consideration, this paper adopts a commonly
used model {\color{black}originally} proposed in \cite{al2014letter}, where the air-to-ground channel can either be {\color{black}an} LoS link
or be an NLoS link. The probabilities of {\color{black}an} LoS and {\color{black}an} NLoS link are denoted by $P_L(\phi)$
and $P_N(\phi)$ and are given by
\begin{align}\label{P_lOS}
 &P_{L}(r_0)=\frac{1}{1+C\exp{(-B(\phi(r_0,H)-C))}},\\\notag
 &P_{N}(r_0)=1-P_{L}(\phi(r_0,H)),
\end{align}
where $\phi(r_0,H)=\arctan{\frac{H}{r_0}}$ is the elevation angle from the UE to the UAV,
B and C are constant parameters determined
by the environment.  As can be seen in (\ref{P_lOS}), with a larger elevation angle,
the link is more likely to be {\color{black}an} LoS link.

Furthermore, the air-to-ground channel gain is modeled as
\begin{align}
 h_0=\frac{|g_{0s}|^2}{(\sqrt{H^2+r_0^2})^{\alpha_s}},
\end{align}
where $s \in \{L,N\}$, $L$ denotes {\color{black}an} LoS link and $N$ denotes an NLoS link,
$g_{0s}$ is the small scale fading {\color{black}channel gain} and obeys Nakagami-m fading with parameter $m_s$,
and $\alpha_s$ is the large scale path loss exponent. Particularly, Rayleigh fading is
assumed for NLoS links, i.e., $m_N=1$.

The ground-to-ground channel between the $i$-th ground BS and the UE is modeled as an
NLoS link. The channel gain is
 $h_i=\frac{|g_i|^2}{r_i^{\alpha_N}}$, where $g_i$ is the {\color{black}small scale Rayleigh fading}
 and $\alpha_N$ is the large scale path loss exponent.

\subsection{Transmission Scheme}
This paper proposes a {\color{black}user-centric cooperative} scheme, which means that the UE in disc $\mathcal{D}$ can be served  either by the UAV only,
the nearest ground BS only, or both the UAV and the nearest ground BS,
depending on {\color{black}the user's connections to the UAV and the nearest ground BS}.
Thus, there are three types of UEs in disc $\mathcal{D}$, denoted by $\mathcal{A}_1$ (nearest ground BS only), $\mathcal{A}_2$ (both the UAV
and the nearest ground BS) and  $\mathcal{A}_3$ (UAV only).
\begin{figure}[!t]
\centering
\includegraphics[width=6in]{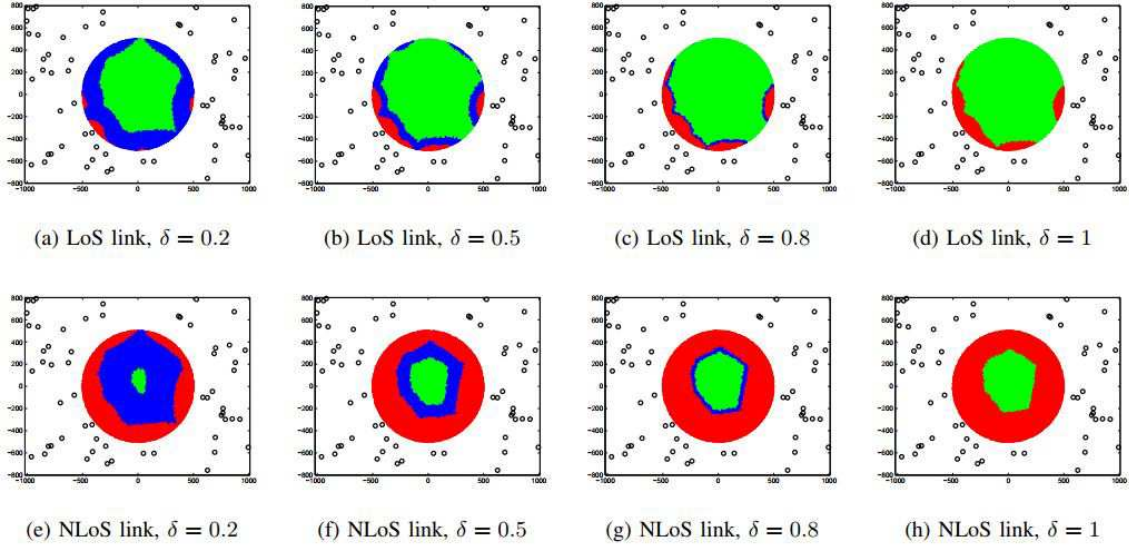}
\caption{Illustration of the user region. $R_c=500$ m , $H=300$ m, $\alpha_L=2.5$,
$\alpha_N=3$, $\lambda=2\times10^{-5}/m^2$. Black circles denote ground BSs generated
from a realization of $\Phi_o$. Red, blue and green regions denote the region of
$\mathcal{A}_1$, $\mathcal{A}_2$ and $\mathcal{A}_3$, respectively. (a)-(d) are the
cases where the air-to-ground links are LoS only and (e)-(f) are the cases where the
air-to-ground links are NLoS only.}
\label{region}
\end{figure}
Mathematically, the UE belongs to which type is determined as follows:
{\color{black}\begin{align}\label{scheme}
\begin{cases}
 \text{UE}\in\mathcal{A}_1,&\text{if } r_1^{\alpha_N} \leq \delta(\sqrt{H^2+r_0^2})^{\alpha_s},\\
 \text{UE}\in\mathcal{A}_2,&\text{if }
 \delta(\sqrt{H^2+r_0^2})^{\alpha_s} < r_1^{\alpha_N} \leq
 \frac{1}{\delta}(\sqrt{H^2+r_0^2})^{\alpha_s},\\
 \text{UE}\in\mathcal{A}_3,&\text{if }
r_1^{\alpha_N} > \frac{1}{\delta}(\sqrt{H^2+r_0^2})^{\alpha_s}.
\end{cases}.
\end{align}}
Note that, the parameter $\delta$ ($0\leq\delta\leq1$) is {\color{black}termed} the {\color{black}cooperation indication} parameter,
which determines the cooperation level between the UAV and the nearest BS.
For example, when $\delta=0$, all the UEs in disc $\mathcal{D}$ belong to
$\mathcal{A}_2$, which means that the UEs are served cooperatively by the UAV and the nearest ground BS;
when $\delta=1$, a UE may possibly belong to $\mathcal{A}_1$ or $\mathcal{A}_3$,
and there is no UE belonging to $\mathcal{A}_3$.   As shown in
Fig. \ref{region}, the cooperation region $\mathcal{A}_3$ decreases with $\delta$.
Another observation from Fig. \ref{region} is that, the region of $\mathcal{A}_3$ is
much smaller when the air-to-ground links are NLoS, which reveals the importance of
the application of the proposed scheme.

This paper only considers {\color{black}the} interference-limited scenario, where the noise {\color{black}are}
omitted compared to the aggregated interference.

When the $\text{UE}\in \mathcal{A}_1$, the UE is served by the nearest ground BS only {\color{black}and} the SIR to decode the UE's message {\color{black}is given by}
\begin{align}
\text{SIR}_1=\frac{h_1}{h_0+\sum\limits_{x_i\in\Phi_o\backslash x_1}h_i}.
\end{align}

When the $\text{UE}\in \mathcal{A}_2$, the UE is served by both the UAV and the nearest ground BS.
Particularly,  this paper considers distributed transmit beamforming at the UAV and the
nearest BS. Consequently, the SIR to decode the UE's message {\color{black}is given by}
\begin{align}
\text{SIR}_2=\frac{(h_0+h_1)}{\sum\limits_{x_i\in\Phi_o\backslash x_1}h_i}.
\end{align}

When the $\text{UE}\in \mathcal{A}_3$, the UE is served by the UAV only, the SIR to
decode the UE's message {\color{black}is given by}
\begin{align}
\text{SIR}_3=\frac{h_0}{\sum\limits_{x_i\in\Phi_o}h_i}.
\end{align}
\section{Performance Analysis}
In this section, we will use the coverage probability
as the criterion to evaluate the performance of the proposed
scheme. The coverage probability is defined as the probability of the event that the
SIR is higher than a threshold $\epsilon$.
The NSE will also be given to reveal the trade off between the system throughput and {\color{black}the number of serving BSs}.
 To evaluate the coverage probability and NSE achieved by the
UE, it is necessary to first obtain the following preliminary results.
\subsection{Distance distribution of the nearest ground BS}
The distribution of the distance from a typical UE to its nearest BS in a standard HPPP model
can be easily obtained and briefly represented {\color{black}\cite{haenggi2012stochastic}}. However, it is much more complicate in the considered
scenario in this paper. The main difficulty in our considered scenario
is caused by the constraint that the ground
BSs should locate outside disc $\mathcal{D}$.
Through rigorous derivations , the following lemma is obtained.
\begin{Lemma}
The conditional pdf of $r_1$ given $r_0$ is given by:
\begin{align}
f_{r_1|r_0}(r)=\begin{cases}0, &r \leq R_c-r_0\\
\lambda\zeta_1(r)e^{-\lambda\zeta_2(r)},& R_c-r_0<r<R_c+r_0\\
2\pi\lambda r e^{-\lambda\left(\pi r^2-\pi R_c^2\right)},&\text{otherwise}
\end{cases},
\end{align}
and the conditional CDF of $r_1$ given $r_0$ is given by:
\begin{align}\label{r1_CDF}
F_{r_1|r_0}(r)=\begin{cases}0, &r \leq R_c-r_0\\
1-e^{-\lambda \zeta_2(r)},& R_c-r_0<r<R_c+r_0\\
1-e^{-\lambda\left(\pi r^2-\pi R_c^2\right)},&\text{otherwise}
\end{cases},
\end{align}
where
\begin{align}
\zeta_1(r)=&2\pi r+
           \frac{r}{r_0}\sqrt{-\frac{(r-r_0-R_c)(r+r_0-R_c)(r-r_0+R_c)(r+r_0+R_c)}{r_0^2}}\\\notag
           &-\frac{r}{r_0}\sqrt{-\frac{(-r+r_0-R_c)(r+r_0-R_c)(-r+r_0+R_c)(r+r_0+R_c)}{r_0^2}}\\\notag
           &-2r\sec^{-1}\frac{2rr_0}{r^2+r_0^2-R_c^2},
\end{align}
and
\begin{align}
\zeta_2(r)=\pi r^2-\theta_1(r)R_c^2+R_c^2\sin\theta_1(r)\cos\theta_1(r)-\theta_2(r)r^2
            +r^2\sin\theta_2(r)\cos\theta_2(r),
\end{align}
$\theta_1(r)=\arccos\frac{R_c^2+r_0^2-r^2}{2R_c r_0}$ and
$\theta_2(r)=\arccos\frac{r_0^2+r^2-R_c^2}{2 r_0 r}$.
\end{Lemma}
\begin{IEEEproof}
Please refer to Appendix A.
\end{IEEEproof}
\subsection{Laplace transform of the interference}
Define $I_2=\sum\limits_{x_i\in\Phi_o\backslash x_1}h_i$, which is the aggregated
interference from the ground BSs farther than the nearest ground BS.
This subsection will focus on calculating the Laplace transform of $I_2$, when $r_0$ and $r_1$ are {\color{black}are assumed to be fixed}. There are conditions
need to be considered which complicate the calculation. One is that the distance from
 the UE to each interfering ground BS which contributes to $I_2$ should {\color{black}be larger} than the
distance from the UE to the nearest ground BS, i.e, $r_i>r_1, i\geq2$.
The other is that each interfering ground BS should {\color{black}locate} outside disc $\mathcal{D}$.
By noting that the calculation will be different for the two cases:
i) $R_c-r_0<r_1<R_c+r_0$, ii) $r_1 \geq R_c+r_0$, the following two lemmas are obtained.
\begin{Lemma}
Define the conditional Laplace transform of $I_2$ when $r_0$ and $r_1$ are fixed
as $\mathcal{L}_{I_2|r_0,r_1}(s)=\mathbb{E}\{\text{exp}(-sI_2)\}$, then
$\mathcal{L}_{I_2|r_0,r_1}(s)$ is given by:
\begin{align}
\mathcal{L}_{I_2|r_0,r_1}(s)=\exp(\eta(s)),
\end{align}
where $\eta(s)$ can be expressed as the following two cases:
\begin{itemize}
  \item when $R_c-r_0<r_1<R_c+r_0$,
   \begin{align}
         \eta(s)\approx&-\frac{2\lambda(\pi-\Theta)s^{\frac{2}{\alpha_N}}}{\alpha_N}
           \bar{B}\left(\frac{1}{1+sr_1^{-\alpha_N}};\frac{2}{\alpha_N},1-\frac{2}{\alpha_N}\right)
           \\\notag
            &-\frac{\lambda\Theta\pi s^{\frac{2}{\alpha_N}}}{N\alpha_N}\sum_{n=1}^{N}
          \sqrt{1-\theta_n^2}\bar{B}\left(\frac{1}{1+s(z(c_n))^{-\alpha_N}};\frac{2}{\alpha_N},1-\frac{2}{\alpha_N}\right),
   \end{align}
  \item when $r_1 \geq R_c+r_0$,
  \begin{align}
   \eta(s)=&-\frac{2\lambda\pi s^{\frac{2}{\alpha_N}}}{\alpha_N}
           \bar{B}\left(\frac{1}{1+sr_1^{-\alpha_N}};\frac{2}{\alpha_N},1-\frac{2}{\alpha_N}\right),
  \end{align}
\end{itemize}
where $\bar{B}(x;a,b)$ is the upper incomplete beta function given by $\bar{B}=\int_x^1 t^{a-1}(1-t)^{b-1}\,dt$,
$\Theta=\arccos{\frac{r_0^2+r_1^2-R_c^2}{2r_0r_1}}$,
$z(\theta)=\sqrt{R_c^2-r_0^2\sin^2\theta}-r_0\cos{\theta}$,
$N$ denotes the parameter for Chebyshev-Gauss quadrature,
$\theta_n=\cos{\frac{(2n-1)\pi}{2N}}$ and $c_n=\frac{\Theta}{2}(\theta_n-1)+\pi$.
\end{Lemma}
\begin{IEEEproof}
Please refer to Appendix B.
\end{IEEEproof}

\begin{Lemma}
The $k$-th ($k \geq 1$) derivative of the Laplace transform $\mathcal{L}_{I_2|r_0,r_1}(s)$ can be
calculated recursively as follows:
\begin{align}
\mathcal{L}_{I_2|r_0,r_1}^{(k)}(s)=\sum_{l=0}^{k-1} { {k-1}\choose{l}} \eta^{(k-l)}(s)
                               \mathcal{L}_{I_2|r_0,r_1}^{(l)}(s),
\end{align}
where $\eta^{(t)}(s)$ is the $t$-th ($t \geq 1 $) derivative of $\eta(s)$, which can be evaluated as follows:
\begin{itemize}
  \item when $R_c-r_0<r_1<R_c+r_0$,
   \begin{align}\label{Lap_d1}
         \eta^{(t)}(s)\approx&t!(-1)^t\lambda\left(\frac{2(\pi-\Theta)s^{\frac{2}{\alpha_N}-t}}{\alpha_N}
           \bar{B}\left(\frac{1}{1+sr_1^{-\alpha_N}};\frac{2}{\alpha_N}+1,t-\frac{2}{\alpha_N}\right)
           \right.\\\notag
            &\left.+\frac{\Theta\pi s^{\frac{2}{\alpha_N}-t}}{N\alpha_N}\sum_{n=1}^{N}
          \sqrt{1-\theta_n^2}\bar{B}\left(\frac{1}{1+s(z(c_n))^{-\alpha_N}};\frac{2}{\alpha_N}+1,t-\frac{2}{\alpha_N}\right)\right),
   \end{align}
  \item when $r_1 \geq R_c+r_0$,
  \begin{align}\label{Lap_d2}
   \eta^{(t)}(s)=&\frac{t!(-1)^t\lambda2\pi s^{\frac{2}{\alpha_N}-t}}{\alpha_N}
           \bar{B}\left(\frac{1}{1+sr_1^{-\alpha_N}};\frac{2}{\alpha_N}+1,t-\frac{2}{\alpha_N}\right).
  \end{align}
\end{itemize}
\end{Lemma}
\begin{IEEEproof}
Please refer to Appendix B.
\end{IEEEproof}
\subsection{Area fraction and coverage probabilities}
An interesting problem is that
what fraction of users in disc $\mathcal{D}$ are served
by different transmission strategies.
To answer this question, the following proposition which provides the
expected area of $\mathcal{A}_1$, $\mathcal{A}_2$ and $\mathcal{A}_3$ in disc $D$ is
first highlighted as follows.
\begin{Proposition}
The expected area of $\mathcal{A}_1$, $\mathcal{A}_2$ and $\mathcal{A}_3$ in disc $D$ can be expressed respectively
as follows:
\begin{align}\label{C_A1}
C_{\mathcal{A}_1}=2\pi\sum_{s\in\{L,N\}}\int_0^{R_c}
                    P_s(r_0)F_{r_1|r_0}(A_s(r_0))r_0\,dr_0,
\end{align}
\begin{align}
C_{\mathcal{A}_2}=2\pi\sum_{s\in\{L,N\}}\int_0^{R_c}
                    P_s(r_0)\left(F_{r_1|r_0}(B_s(r_0))-F_{r_0|r_1}(A_s(r_0))\right)r_0\,dr_0,
\end{align}
\begin{align}
C_{\mathcal{A}_3}=2\pi\sum_{s\in\{L,N\}}\int_0^{R_c}
                     P_s(r_0)\left(1-F_{r_1|r_0}(B_s(r_0))\right)r_0\,dr_0,
\end{align}
where $A_s(r_0)=\left(\delta(\sqrt{H^2+r_0^2})^{\alpha_s}\right)^{\frac{1}{\alpha_N}}$ and
$B_s(r_0)=\left(\frac{1}{\delta}(\sqrt{H^2+r_0^2})^{\alpha_s}\right)^{\frac{1}{\alpha_N}}$,
$s\in\{L,N\}$.
\end{Proposition}
\begin{IEEEproof}
Please refer to Appendix C.
\end{IEEEproof}
With Proposition 1,  the area fraction can be defined as:
\begin{align}\label{E_area_fraction}
 \bar{C}_{\mathcal{A}_i}=\frac{C_{\mathcal{A}_i}}{\pi R_c^2},i\in \{1,2,3\},
\end{align}
which is the expected area of $C_{\mathcal{A}_i}$ normalized by the area of disc $\mathcal{D}$.
Note that, the area fraction is affected by many parameters, such as $\delta$,
$\lambda$, etc. Unfortunately, the impact of these parameters cannot be captured straightforwardly
due to the complex expression of $C_{\mathcal{A}_i}$. Even so, with the help of the proof
as shown in Appendix C,  some insights are obtained as highlighted in
the following corollaries.
\begin{Corollary}
With $0<\delta<1$, $\bar{C}_{\mathcal{A}_1}$ and $\bar{C}_{\mathcal{A}_3}$
{\color{black}increase} with $\delta$, while $\bar{C}_{\mathcal{A}_2}$ decreases with $\delta$ .
\end{Corollary}
\begin{Corollary}
With $0<\delta<1$, $\bar{C}_{\mathcal{A}_1}$ increases with $\lambda$
and $\bar{C}_{\mathcal{A}_3}$ decreases with $\lambda$.
\end{Corollary}

\begin{Remark}
The impact of $R_c$ and $H$ on $C_{\mathcal{A}_i}$ is {\color{black}difficult} to be obtained.
For example, when $H$ increases, it can be seen from
(\ref{P_lOS}) that the probability $P_L(r_0)$ increases while  $P_N(r_0)$ shows the
opposite trend. Besides, both $A_s(r_0)$ and $B_s(r_0)$ {\color{black}increase} with H. Thus,
 it is not easy to evaluate the impact of $H$ when considering all these factors. The impact
 of $R_c$ and $H$ will {\color{black}evaluated by using} numerical results.
\end{Remark}

With the help of Lemma $2$ and Lemma $3$,  we have the following three lemmas
which characterize the conditional coverage probability
given $r_0$ and $r_1$ achieved by the UE, when the UE belongs to $\mathcal{A}_1$,
$\mathcal{A}_2$ and $\mathcal{A}_3$, respectively.

\begin{Lemma}\label{P_case1}
When $\text{UE} \in \mathcal{A}_1$, the conditional coverage probability achieved by the UE
given $r_0$ and $r_1$ can be expressed as follows:
\begin{align}
P^1(r_0,r_1)=\frac{\mathcal{L}_{I_2|r_0,r_1}(r_1^{\alpha_N}\epsilon)}{\left(1+\frac{r_1^{\alpha_N}\epsilon}{(H^2+r_0^2)^{\frac{\alpha_L}{2}}m_L}\right)^{m_L}}
              .
\end{align}
\end{Lemma}
\begin{IEEEproof}
Please refer to Appendix D.
\end{IEEEproof}

Consider a special case when there's no UAV employed in disc $D$ to address
the emergency and the UEs in disc $\mathcal{D}$ are only served by the nearest BS
outside the disc. In this case, the performance of the UE can {\color{black}be} easily obtained from the
proof of Proposition \ref{P_case1}, which is highlighted as follows.

\begin{Corollary}
When there is no UAV and the UE is only served by the nearest BS, the conditional coverage probability achieved by the UE
given $r_0$ and $r_1$ can be expressed as follows:
\begin{align}
\tilde{P}^1(r_0,r_1)=\mathcal{L}_{I_2|r_0,r_1}(r_1^{\alpha_N}\epsilon).
\end{align}
\end{Corollary}

\begin{Lemma}
When $\text{UE} \in \mathcal{A}_2$, the conditional coverage probability achieved by the UE
given $r_0$ and $r_1$ can be expressed as follows:
\begin{align}
P^2(r_0,r_1)=\sum_{j=0}^{1}\sum_{k=1}^{\alpha_j}\frac{A_{jk}}{\beta_j^k}\sum_{l=0}^{k-1}
              \frac{(-u_j)^l}{l!}\mathcal{L}_{I_2|r_0,r_1}^{(l)}(u_j)
,\end{align}
\end{Lemma}
where $\alpha_0=m_L$, $\beta_0=m_L(H^2+r_0^2)^{\frac{\alpha_L}{2}}$,
$\alpha_1=1$, $\beta_1=r_1^{\alpha_N}$,
$u_j=\beta_j\epsilon$, and
\begin{align}
A_{jk}=(-1)^{\alpha_j-k}\frac{\beta_0^{\alpha_0}\beta_1^{\alpha_1}
           (\alpha_{1-j}+\alpha_j-k-1)!}{(\alpha_j-k)!(\alpha_{1-j}-1)!}
           (\beta_{1-j}-\beta_j)^{-\alpha_{1-j}-\alpha_j+k} .
\end{align}
\begin{IEEEproof}
Please refer to Appendix D.
\end{IEEEproof}

\begin{Lemma}
When $\text{UE} \in \mathcal{A}_3$, the conditional coverage probability achieved by the UE
given $r_0$ and $r_1$ can be expressed as follows:
\begin{align}
P^3(r_0,r_1)=\sum_{l=0}^{m_L-1}\frac{(-u)^l}{l!}\mathcal{L}_{h_1+I_2|r_0,r_1}^{(l)}(u),
\end{align}
where $u=m_L (H^2+r_0^2)^{\frac{\alpha_L}{2}}\epsilon$, $\mathcal{L}_{h_1+I_2|r_0,r_1}^{(l)}(u)$ is the
$l$-th derivative of the Laplace transform for $h_1+I_2$, which is given by:
\begin{align}
\mathcal{L}_{h_1+I_2|r_0,r_1}^{(l)}(u)=\sum_{p=0}^{l} {l \choose p}
             \mathcal{L}_{I_2|r_0,r_1}^{(p)}(u)\mathcal{L}_{h_1|r_0,r_1}^{(l-p)}(u),
\end{align}
and
\begin{align}
 \mathcal{L}_{h_1|r_0,r_1}^{(l-p)}(u)=\frac{r_1^{\alpha_N}t!(-1)^t}{\left(u+r_1^{\alpha_N}\right)^{t+1}}.
\end{align}
\end{Lemma}
\begin{IEEEproof}
Please refer to Appendix D.
\end{IEEEproof}

Based on Lemma $1$ and Lemmas $4$-$6$, by taking expectation with respect to $r_1$,
the conditional probability given $r_0$ can be obtained as shown in the following theorem.
\begin{theorem}\label{theorem_r0}
The conditional coverage probability achieved by the UE given $r_0$ can be calculated as follows:
\begin{align}
 P(r_0)=\sum_{i=1}^{3}Pc_{i}(r_0).
\end{align}
where $Pc_{i}(r_0)$ is the conditional probability given $r_0$ for the event that the UE belongs to
$\mathcal{A}_i$ and the QoS is satisfied, $Pc_{i}(r_0)$ can be expressed as follows:
\begin{align}
Pc_{1}(r_0)&=\int_0^{A(r_0)}P^1(r_0,r_1)f_{r_1|r_0}(r_1)\,dr_1,\\\notag
Pc_{2}(r_0)&=\int_{A(r_0)}^{B(r_0)}P^2(r_0,r_1)f_{r_1|r_0}(r_1)\,dr_1,\\\notag
 Pc_{3}(r_0)&=\int_{B(r_0)}^{\infty}P^3(r_0,r_1)f_{r_1|r_0}(r_1)\,dr_1.
\end{align}
\end{theorem}

\subsection{Normalized spectral efficiency}
{\color{black}For the sake of the system throughput, it is better to let all UEs reside in $\mathcal{A}_2$.
However, this is at the expense of occupying more BSs (both the UAV and the nearest ground BS is occupied), compared to serving UEs by the UAV only or the nearest ground BS only. In order to consider the trade-off between the system throughput and the number of serving BSs,} the
normalized spectral efficiency (NSE) of the malfunction area is used in this paper which is defined as follows:
\begin{align}
  \text{NSE}=\sum_{i=1}^{3}\frac{Pc_{i}\log(1+\epsilon)}{N_i},
\end{align}
where $Pc_i$ is the probability of the event that the UE belongs to $\mathcal{A}_i$ and the
rate is guaranteed and is given by $Pc_i=\int_{0}^{R_c}Pc_i(r_0)\frac{2r_0}{R_c^2}\,dr_0$, and
$N_i$ is the number of BSs used in the transmission scheme for UEs in $\mathcal{A}_i$, i.e.,
$N_1=1$, $N_2=2$, and $N_3=1$.

\section{Numerical Results}
In this section, numerical results are presented to demonstrate the performance of the proposed
scheme and also verify the developed analytical results. Unless {\color{black}stated otherwise}, the parameters
are set as follows: $B=0.136$, $C=11.95$, $R_c=500$ m, $H=300$ m, $\alpha_L=3=2.5$, $\alpha_N=3$,
$m_L=4$.
\begin{figure*}[!t]
\centering
\subfloat[Area fraction vs. $H$, $R_c=500$ m]{\includegraphics[width=3.2in]{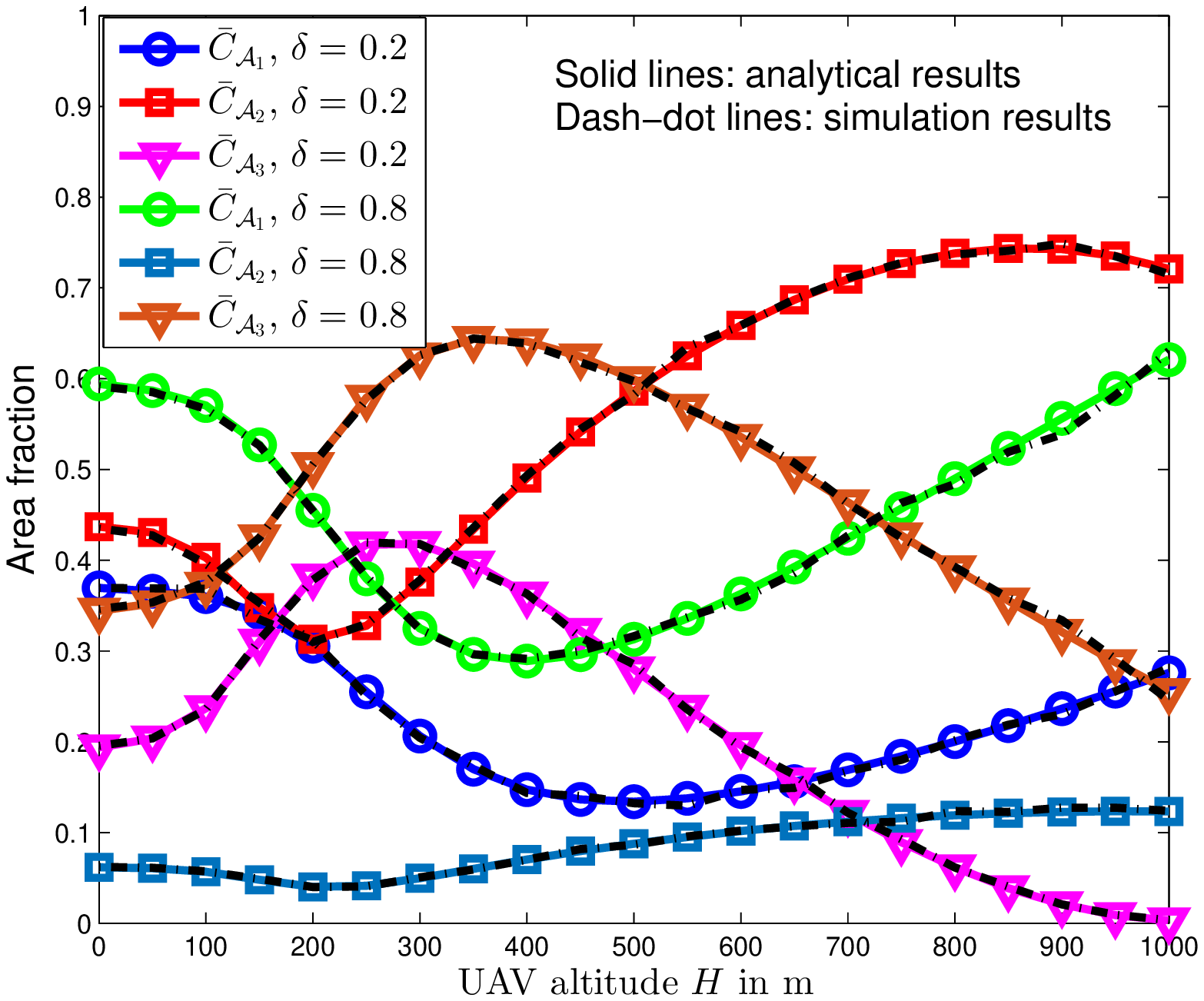}%
\label{compare1}}
\hfil
\subfloat[Area fraction vs. $R_c$, $H=300$ m]{\includegraphics[width=3.2in]{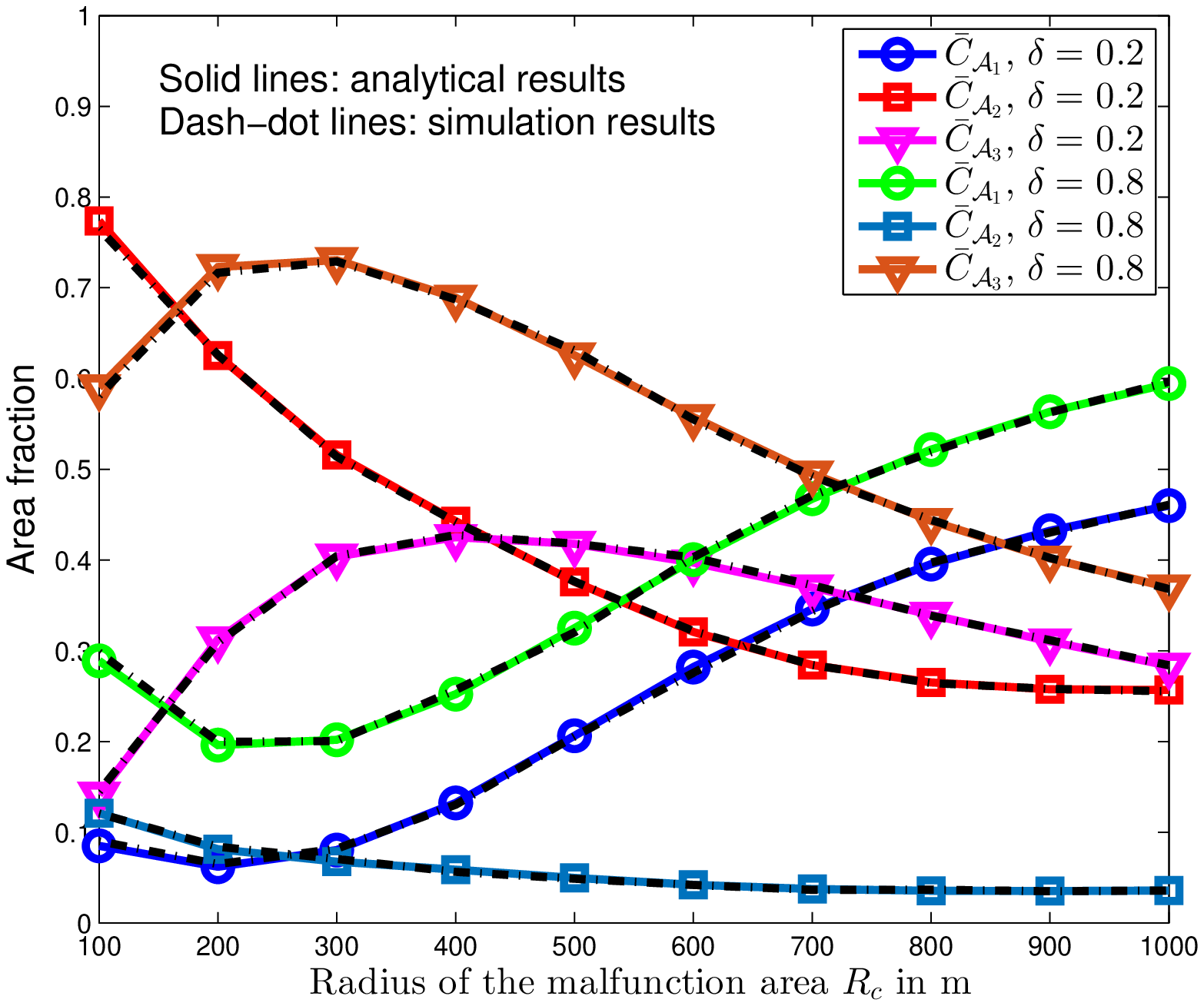}%
\label{compare2}}
\caption{Impact of $R_c$ and $H$ on $\bar{C}_{\mathcal{A}_i}$, $i=1,2,3$.  }
\label{Area_fraction}
\end{figure*}

Fig. \ref{Area_fraction}(a) and Fig. \ref{Area_fraction}(b) {\color{black}show} how the area fraction of $\mathcal{A}_1$,  $\mathcal{A}_2$ and  $\mathcal{A}_3$ varies with the UAV altitude $H$ and the malfunction area $R_c$, respectively.
In both the figures, simulation results perfectly match the theoretic results based in
(\ref{E_area_fraction}), which verifies the developed analytical results.
{\color{black}From both Fig. \ref{Area_fraction}(a) and Fig. \ref{Area_fraction}(b), it is observed that $\bar{C}_{\mathcal{A}_1}$ and $\bar{C}_{\mathcal{A}_3}$ with $\delta=0.2$ are smaller than that with $\delta=0.8$. In the contrary,
$\bar{C}_{\mathcal{A}_2}$ shows the opposite trend. These observations are consistent with the conclusion as highlighted in Corollary 1.}
Fig. \ref{Area_fraction}(a) shows that: when $H$ varies from $0$ to $1000$ m, a) $\bar{C}_{\mathcal{A}_1}$ and  $\bar{C}_{\mathcal{A}_2}$
first decrease with $H$ and then increase; b) $\bar{C}_{\mathcal{A}_3}$  first increases with $H$ and then decreases.
Fig. \ref{Area_fraction}(b) shows that: when $R_c$ varies from $100$ to $1000$ m,
a) $\bar{C}_{\mathcal{A}_1}$ first slightly decreases with $R_c$ and then increases;
b) $\bar{C}_{\mathcal{A}_2}$ decreases with $R_c$;
c) $\bar{C}_{\mathcal{A}_3}$ first increases with $R_c$ and then decreases.
\begin{figure}[!t]
\centering
\includegraphics[width=3.5in]{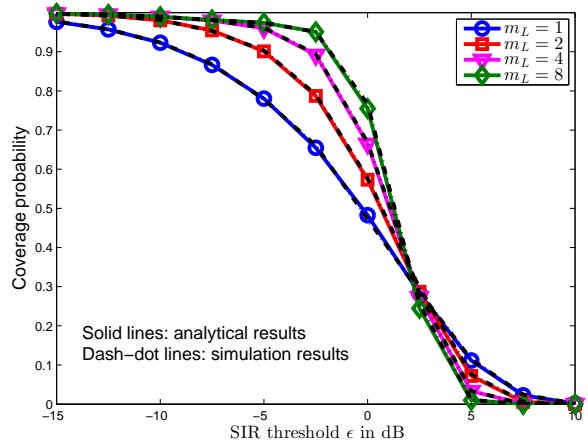}
\caption{Coverage probability $P(r_0)$ vs. SIR threshold $\epsilon$ in dB. $r_0=200$ m, $\delta=0.2$.}
\label{accuracy}
\end{figure}

Fig. \ref{accuracy} shows the coverage probabilities achieved by a UE which locates at {\color{black}a} fixed distance from the origin in the proposed scheme.
The analytical results are based on Theorem \ref{theorem_r0}.
The simulation results are obtained by using Monte Carlo simulations. Specifically, we do $20000$ independent drops of points in a large circular simulation area with radius $40$ km, for each point shown in Fig. \ref{accuracy}.
It is shown in Fig. \ref{accuracy} that the simulation results perfectly match the theoretical results, which verifies the accuracy of the developed analysis.
\begin{figure}[!t]
\centering
\includegraphics[width=3.5in]{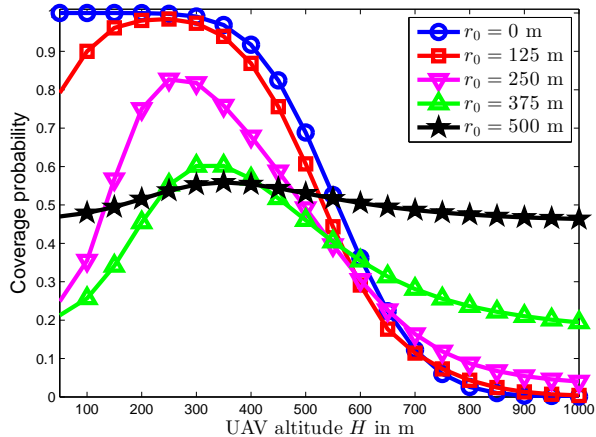}
\caption{Coverage probability $P(r_0)$ vs. UAV altitude $H$ in m. $\delta=0.2$, $\epsilon=0.5$.}
\label{impact_H}
\end{figure}

Fig. \ref{impact_H} shows the coverage probabilities versus UAV altitude $H$, achieved by UEs with different locations. Note that, the UAV altitude $H$ has dual effects on the air-to-ground channel.
On the one side, as $H$ increases, the elevation angle from the UE to the UAV also increases. {\color{black}As} a result, the probability for an LoS link is enlarged, which {\color{black}has a} positive effect on the propagation gain.
{\color{black}On the other hand}, as $H$ increases, the distance from the UE to the UAV also {\color{black}increases}, which {\color{black}has a} negative effect on the propagation gain due to large scale path losses.
{\color{black}Furthermore}, it is obvious that $H$ also affects the transmission strategy for the UE.

Interestingly, as shown in Fig. \ref{impact_H}, the UAV altitude $H$ has a different impact on coverage probabilities for different UE locations.
For example, when the UE locates at the origin, i.e., $r_0=0$ m, the coverage probability {\color{black}decreases} with $H$. The reason for this phenomenon is that when $r_0=0$, the elevation angle from the UE to the UAV is constantly $90$ degrees and will not change with $H$.  {\color{black}Thus} $H$ has no effect on the probability of an LoS link and only impact the large-scale path loss.

When $r_0=125$ m, $r_0=250$ m, and $r_0=375$ m, the coverage probability first increases with $H$, then decreases, and finally maintains at a pretty low level.  Because in {\color{black}at} low altitude, increasing $H$ results in {\color{black}a} rapid increase in LoS probability, which will dramatically improve the air-to-ground link.
While {\color{black}at high altitude}, the link from the UE to the UAV is almost sure to be an LoS link and $H$
only {\color{black}affects}  the distance as well as the transmission strategy for the UE.

When the UE locates at the edge of the circular, i.e., $r_0=500$ m, the impact of $H$ can be neglected. For the reason that the UE is almost sure to be served by the  nearest ground BS and the interference
from the UAV is fairly small due to the large distance.

\begin{figure*}[!t]
\centering
\subfloat[Coverage probability]{\includegraphics[width=3.2in]{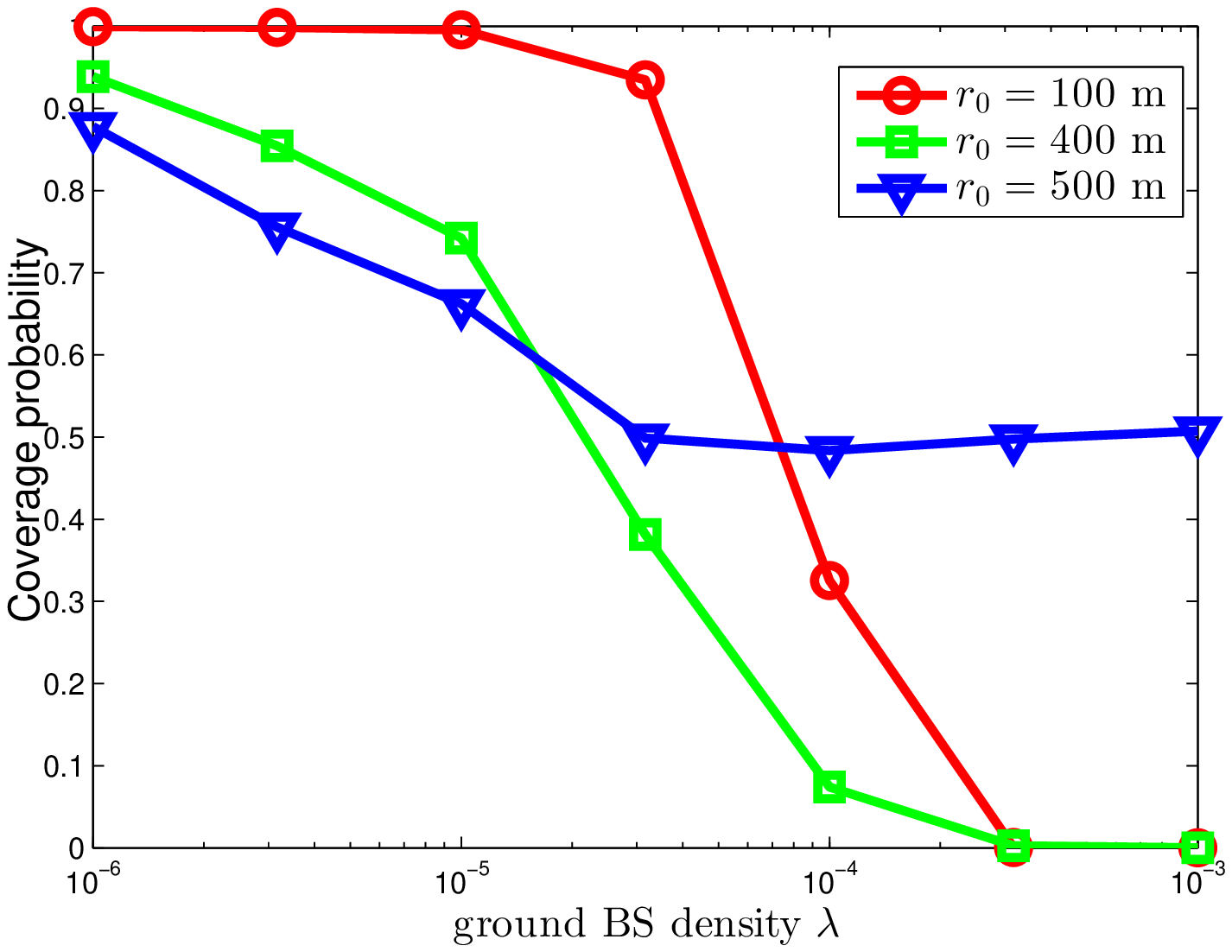}%
\label{impact_lambda1}}
\hfil
\subfloat[Probabilities of different strategies]{\includegraphics[width=3.2in]{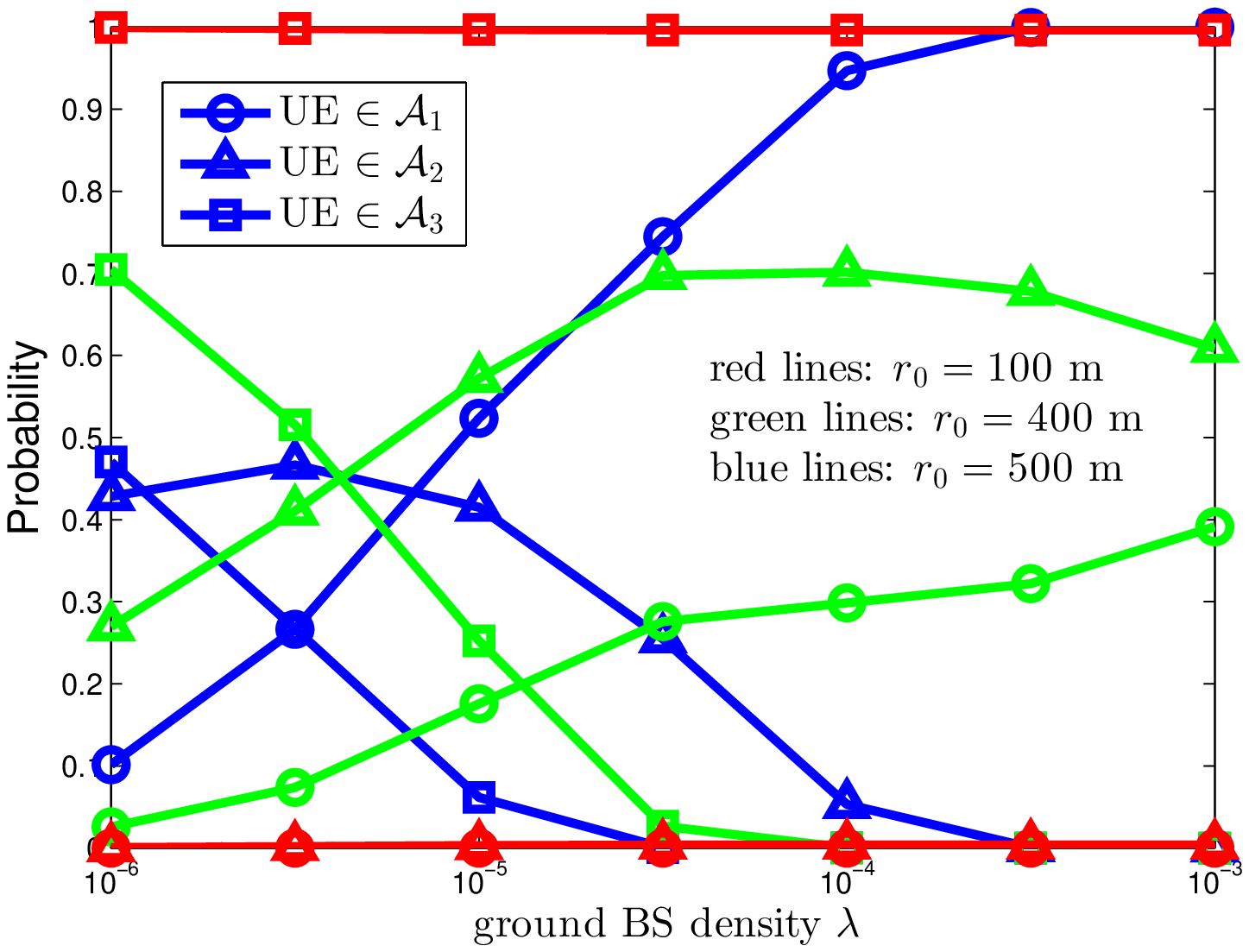}%
\label{impact_lambda2}}
\caption{Impact of ground BS density $\lambda$ on coverage probability. $\delta=0.2$, $\epsilon=0.5$.}
\label{impact_lambda}
\end{figure*}

Fig. \ref{impact_lambda} studies the impact of the ground BS density $\lambda$ on the performance of the proposed scheme. Fig. \ref{impact_lambda}(a) shows the coverage probability versus $\lambda$ for UEs with different locations, {\color{black}and} Fig. \ref{impact_lambda}(b) shows the probabilities of the events that the UE
belongs to $\mathcal{A}_1$, $\mathcal{A}_2$ and $\mathcal{A}_3$.
As shown in the figure, when $r_0=100$ m, the UE is {\color{black}almost always} served by the UAV only, thus the coverage probability deceases with $\lambda$ due to the increased interference from ground BSs.
It can also be seen from Fig. \ref{impact_lambda}, when $r_0=500$ m, the coverage probability first decreases with $\lambda$ and then maintains at about 0.5. This can be explained as follows. As shown in Fig. \ref{impact_lambda}(b), {\color{black}at low $\lambda$}, the UE belongs to $\mathcal{A}_2$  and $\mathcal{A}_3$ with high probability, {\color{black}in this case, the increasing interference from other ground BSs dominates the impact.}
While {\color{black} at high $\lambda$}, the UE is only served by the nearest ground BS due to {\color{black}the} very small distance. {\color{black}In this case, on the one hand, increasing $\lambda$ results in decreasing the distance from the UE to the nearest ground BS which is positive to the coverage. On the other hand, increasing $\lambda$ also results in increasing the interference from other ground BSs which is negative to the coverage. Consequently, the above two kinds of effect cancel each other and hence the coverage probability stays at a steady level.}

\begin{figure*}[!t]
\centering
\subfloat[Coverage probability]{\includegraphics[width=3.2in]{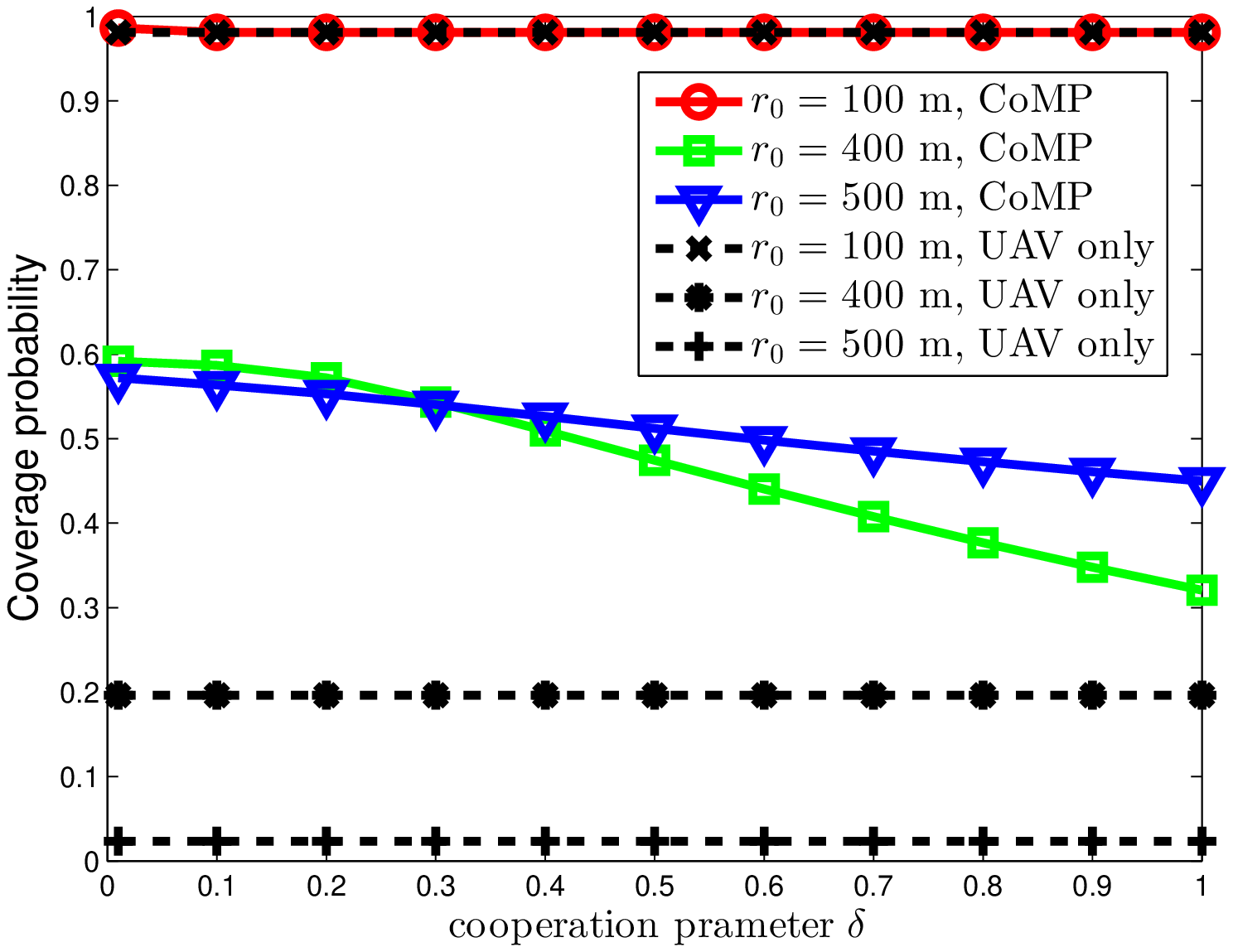}%
\label{impact_delta1}}
\hfil
\subfloat[Probabilities of different strategies]{\includegraphics[width=3.2in]{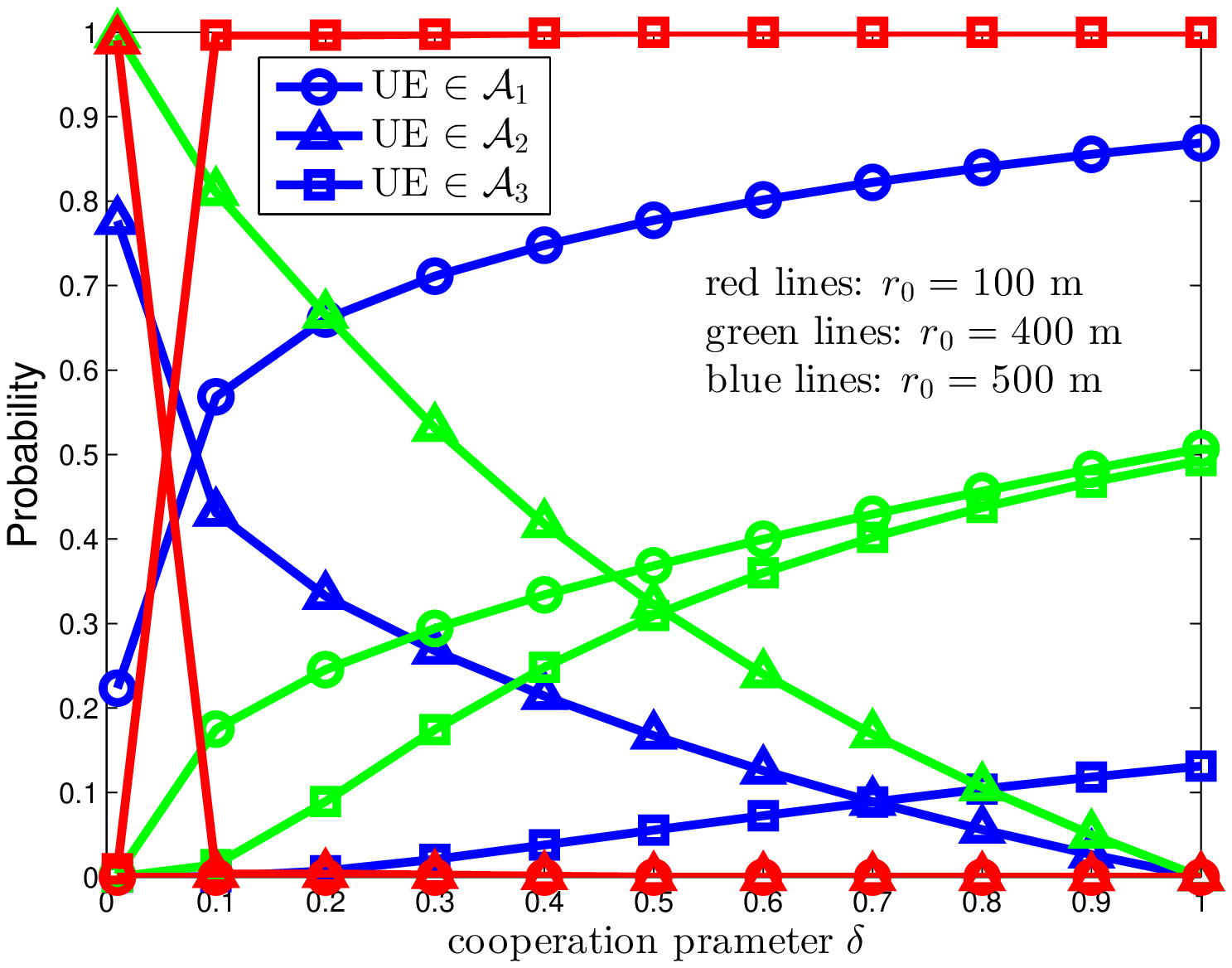}%
\label{impact_delta2}}
\caption{Impact of cooperation parameter $\delta$ on coverage probability. $\epsilon=0.5$.}
\label{impact_delta}
\end{figure*}

Fig. \ref{impact_delta} studies the impact of the cooperation parameter $\delta$ on the performance achieved by the proposed cooperative scheme and the benchmark scheme in \cite{wang2018modeling2} where UEs in the considered scenario are served by the mounted UAV only. From Fig. \ref{impact_delta}, we have the following observations.
When $r_0=100$ m, the cooperation parameter $\delta$ has no effect on the coverage probability of the proposed scheme. Because the UE is almost sure to be served by the UAV only as shown in Fig. \ref{impact_delta}(b). This also explains the fact that the proposed  scheme has the same performance
as the benchmark scheme as shown in Fig. \ref{impact_delta}(a).
When $r_0=400$ m and $r_0=500$ m, the coverage probabilities decrease with $\delta$. For example, with $r_0=400$ m, $P(r_0)$ decreases from $0.6$ to $0.3$. This can be explained from Fig. \ref{impact_delta}(b) that as $\delta$ increases, the probability of the event that the UE belongs to $\mathcal{A}_2$ decreases. {\color{black}As a result, it is more likely that the UE is served by the UAV only or the nearest ground BS only}.
It can also be seen from the figure that the proposed cooperative scheme always outperforms the benchmark scheme in terms of {\color{black}the} coverage probability, even when $\delta=1$ which means the UE can only
be served by a UAV or a nearest ground BS. This is because BS association is {\color{black}carried out} when $\delta=1$ which is ignored in the benchmark scheme.

\begin{figure*}[!t]
\centering
\subfloat[Coverage probability]{\includegraphics[width=3.2in]{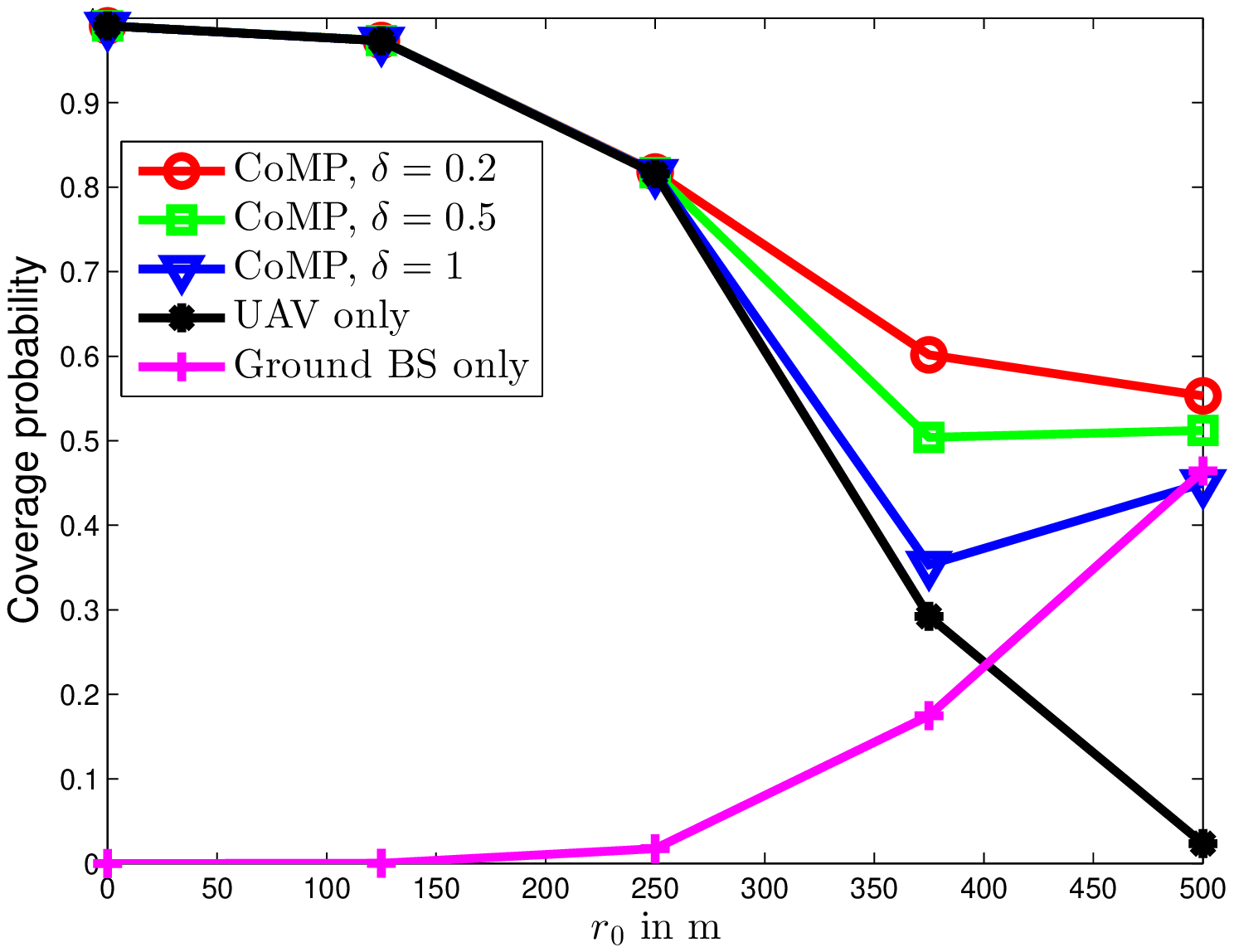}%
\label{compare1}}
\hfil
\subfloat[NSE]{\includegraphics[width=3.2in]{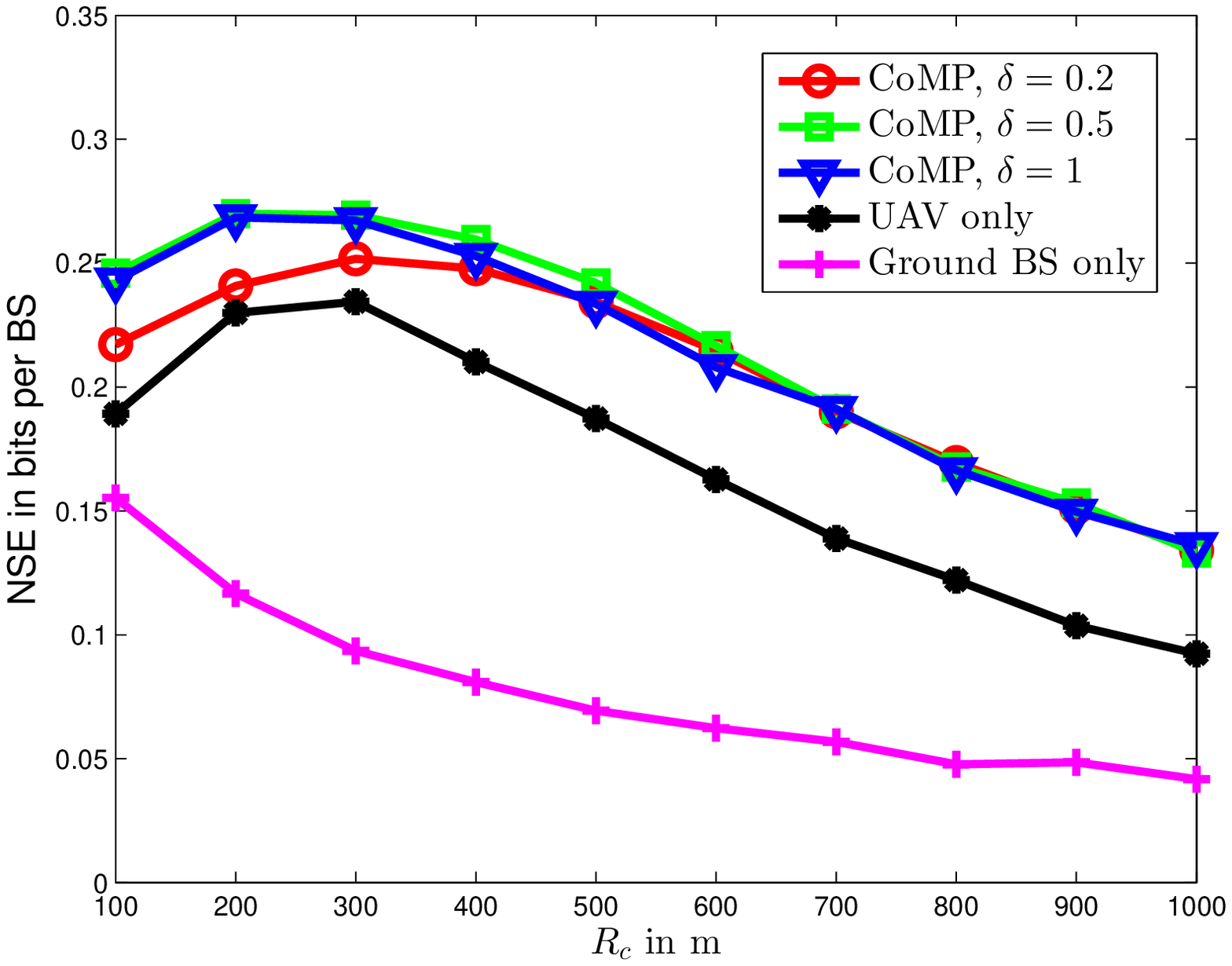}%
\label{compare2}}
\caption{Comparisons between the proposed cooperative scheme and the ``UAV only'' and the ``ground BS only'' scheme. $\epsilon=0.5$.}
\label{compare}
\end{figure*}

Fig. \ref{compare} shows the comparison between the proposed cooperative scheme and two benchmarks.
In the benchmark scheme {\color{black}termed ``UAV only''}, the UE is served by the UAV, while in the scheme {\color{black}termed ``ground BS only''}, it is assumed that there is no UAV employed {\color{black}in} the malfunction area and the UE is served by the nearest ground BS only.
Fig. \ref{compare}(a) shows the coverage probabilities versus the UE location $r_0$. It is shown that
when $r_0$ is small, the proposed scheme achieves similar performances compared to the ``UAV only'' scheme. While as $r_0$ increases, the proposed scheme outperforms the ``UAV only'' scheme.
This can be explained as follows. When $r_0$ is small, the proposed scheme assigns the UE to $\mathcal{A}_3$ with high probability, due to the very {\color{black}small} distance to the UAV and the high probability of an LoS link. However, as $r_0$ increases, the channel between the UAV and the UE becomes
weaker, after realizing this change, the proposed scheme automatically {\color{black}switches} the transmission strategy
according to expression (\ref{scheme}) by assigning the UE to $\mathcal{A}_2$ or $\mathcal{A}_3$,  in order to provide better service compared to the ``UAV only'' scheme.
It is also observed in Fig. \ref{compare}(a) that the proposed scheme significantly outperforms the ``ground BS only'' scheme for most of $r_0$. However, when $r_0$ approaches to $R_c$, i.e., the UE locates at the edge of the malfunction area, the ``ground BS only'' achieves similar performance compared to the proposed scheme.
Fig. \ref{compare}(b) shows the NSE versus the radius of the circular malfunction area $R_c$.
From Fig. \ref{compare}(b), is is shown that the proposed scheme outperforms the ``UAV only'' and
the ``ground BS only'' scheme in terms of NSE. It is also observed that as $R_c$ increases, the NSEs
achieved by the proposed scheme for different $\delta$ are the same.
\section{Conclusions}
In this paper, a {\color{black}user-centric} cooperative scheme has been proposed for a UAV assisted malfunction area which is surrounded by PPP modeled ground BSs.
{\color{black}The} probabilistic LoS/NLoS channel model has been taken into consideration to model the air-to-ground {\color{black}channels}.
Average received power has been used as a criterion to determine which transmission strategy should be
applied to serve the UE, i.e., the UAV only, the nearest ground BS only, or both of them.
A parameter $\delta$ has been introduced to tune the cooperation level of the proposed scheme.
Analytical framework has been developed to evaluate the performance by {\color{black}developing} the expressions for the coverage probability and NSE, which has been verified by computer simulations.
Extensive numerical results have been presented to demonstrate the impact of different parameters on the performance achieved the proposed scheme. It has been shown that the proposed scheme has superior
performance over the {\color{black}``UAV only''} scheme in \cite{wang2018modeling2} and the {\color{black}``ground BS only''} scheme.

Although the superiority of the proposed scheme has been {\color{black}demonstrated} in this paper, {\color{black}there are still some important topics for future research about the application of UAVs to the considered malfunction area.}
For example, whether moving UAVs can provide better performance to such a scenario is still unknown.
{\color{black}Besides, as the size of the malfunction area increases, it is not enough to utilize only one UAV in the malfunction area and it is necessary to deploy multiple UAVs.}
\appendices
\section{Proof for Lemma 1}
\begin{figure*}[!t]
\centering
\subfloat[Case I: $R_c-r_0<r<R_c+r_0$]{\includegraphics[width=3.2in]{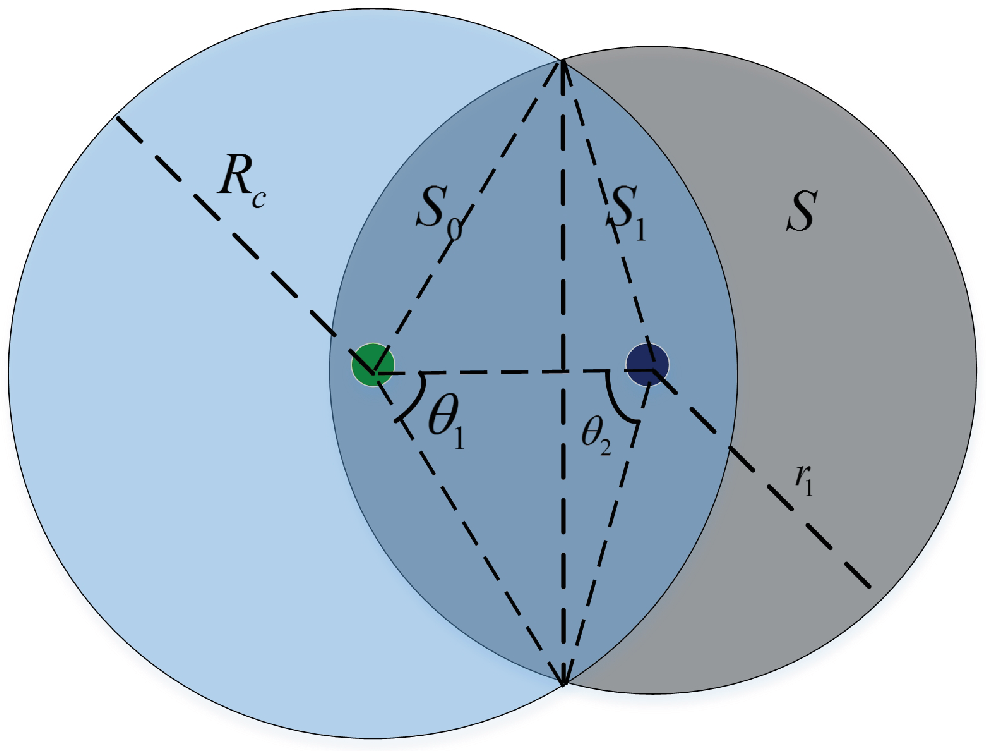}}
\hfil
\subfloat[Case II: $r>R_c+r_0$]{\includegraphics[width=2.8in]{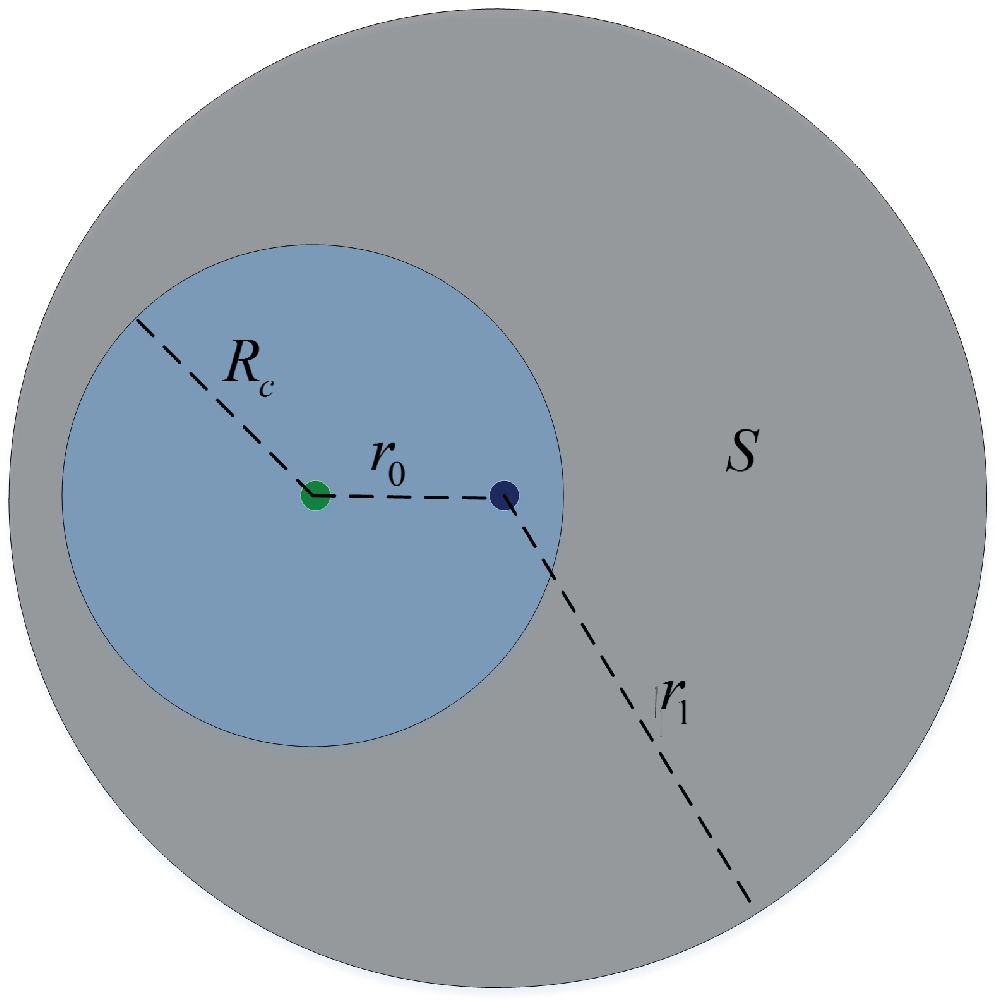}}%
\caption{Illustration of the cases when $R_c-r_0<r<R_c+r_0$ and $r>R_c+r_0$.}
\label{proof_lemma1}
\end{figure*}

Note that, since there's no BS in disc $\mathcal{D}$, the value range of $r_1$ {\color{black}has}
to satisfy $r_1>R_c-r_0$. To obtain the conditional pdf of $r_1$, we need to first calculate the
conditional CDF of $r_1$, which is given by:
\begin{align}{\label{S_r}}
 F_{r_1|r_0}(r)&=1-\text{Pr}\left(r_1>r | r_0\right)\\\notag
               &=1-e^{-\lambda S(r)},
\end{align}
where the last step follows from the fact that the BSs are HPPP distributed outside disc
$\mathcal{D}$. Denote the disc centered at the UE with radius $r_1=r$ by $\mathcal{D}_1$,
then $S(r)$ in (\ref{S_r}) is the area of the region which can be represented by $\mathcal{D}_1-\mathcal{D}_1\cap \mathcal{D}$.

Note that, as shown in Fig. \ref{proof_lemma1}(a) and Fig. \ref{proof_lemma1}(b), the calculations for $S(r)$ when $R_c-r_0<r<R_c+r_0$ and
when $r>R_c+r_0$ are different.

\begin{enumerate}
  \item When $R_c-r_0<r<R_c+r_0$, to calculate $S(r)$, we need to first calculate
  $S_0(r)$ and $S_1(r)$.  With the help of Fig. \ref{proof_lemma1}, it is obtained that $S_0(r)$ can be expressed
as follows:
\begin{align}
 S_0(r)=\theta_1(r)R_c^2-R_c^2\sin{\theta_1(r)}\cos{\theta_1(r)},
\end{align}
where $\theta_1(r)=\arccos\frac{R_c^2+r_0^2-r^2}{2R_c r_0}$. It is worth pointing out
that, when $r$ changes, the value range of $\theta_1(r)$ is $(0,\pi]$.

Similarly, $S_1(r)$ can be expressed as follows:
\begin{align}
S_1(r)=\theta_2(r)r^2-r^2\sin{\theta_2(r)}\cos{\theta_2(r)},
\end{align}
where $\theta_2(r)=\arccos\frac{r_0^2+r^2-R_c^2}{2 r_0 r}$, and when $r$ changes,
the value range of $\theta_1(r)$ is also $(0,\pi]$.

Then $S_(r)$ can be expressed as
\begin{align}
S(r)=\pi r^2-S_0(r)-S_1(r).
\end{align}

\item When $r>R_c+r_0$, $S(r)$ can be easily obtained as follows:
  \begin{align}
   S(r)=\pi r^2-\pi R_c^2.
  \end{align}
\end{enumerate}

Until now, we have obtained the conditional CDF of $r_1$.  By taking the
derivative of $F_{r_1|r_0}(r)$, the conditional pdf of $r_1$ given $r_0$ can be
obtained and the proof for Lemma 1 is complete.

\section{Proof for Lemma 2 and Lemma 3}
\subsection{Proof for Lemma 2}
The Laplace transform of $I_2$ can be calculated as follows:
\begin{align}
\mathcal{L}_{I_2|r_0,r_1}(s)=&\mathbb{E}\left\{\text{exp}(-sI_2)\right\}\\\notag
=&\mathbb{E}\left\{\text{exp}\left(-s\sum\limits_{i=2}^{\infty}h_i\right)\right\}\\\notag
=&\mathbb{E}_{y_i,g_i}\left\{\prod_{i=2}^{\infty}\text{exp}\left(-s\frac{|g_i|^2}{||y_i-x_0||^{\alpha_N}}\right)
                           \right\}\\\notag
{=}&\mathbb{E}_{y_i}\left\{\prod_{i=2}^{\infty} \frac{1}{1+\frac{s}{||y_i-x_0||^{\alpha_N}}}
                       \right\},
\end{align}
where the last step follows from {\color{black}the fact that} the small scale fading gains $|g_i|^2$ are independently  exponential variables with parameter $1$.

By applying the probability generating functional (PGFL) of the {\color{black}HPPP},
$\mathcal{L}_{I_2|r_0,r_1}(s)$ can be further expressed as follows:
\begin{align}
 \mathcal{L}_{I_2|r_0,r_1}(s)=\exp\left(-\lambda
  \underset{\mathcal{R}(r_0,r_1)}{\int}\left(1-\frac{1}{1+\frac{s}{||y-x_0||^{\alpha_N}}}\right)\,dy
  \right)
\end{align}
where $\mathcal{R}(r_0,r_1)$ denotes the integration region which can be determined by
both $r_0$ and $r_1$. Note that, $\mathcal{R}(r_0,r_1)$ can be written as
\begin{align}
 \mathcal{R}(r_0,r_1)=\left\{ y \big| ||y-x_0||>r_1, ||y||>R_c \right\},
\end{align}
where $||y-x_0||>r_1$ means that the distance from the UE to the interfering BS should
be larger than {\color{black}that of} the nearest BS, and $||y||>R_c$ means that the interfering BS should locate
outside disc D.

\begin{figure}[!t]
\centering
\includegraphics[width=3.5in]{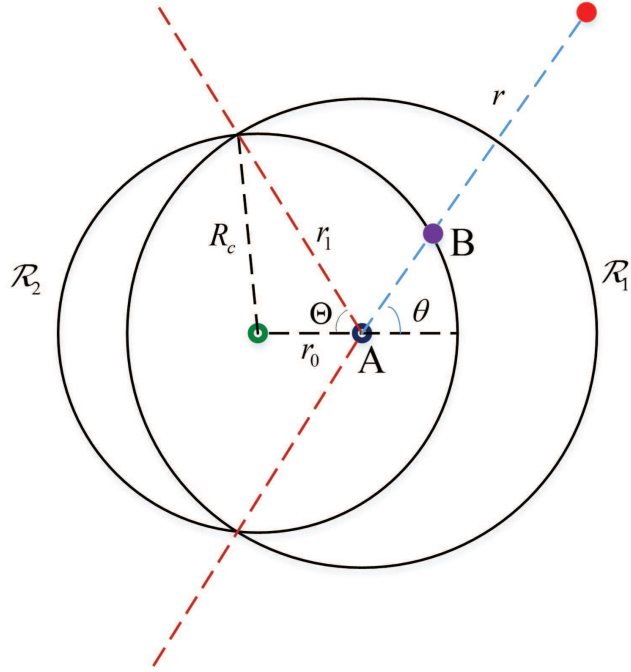}
\caption{Illustration of the system model when $R_c-r_0<r_1<R_c+r_0$}
\label{Proof_Lap_case1}
\end{figure}

\begin{figure}[!t]
\centering
\includegraphics[width=3.5in]{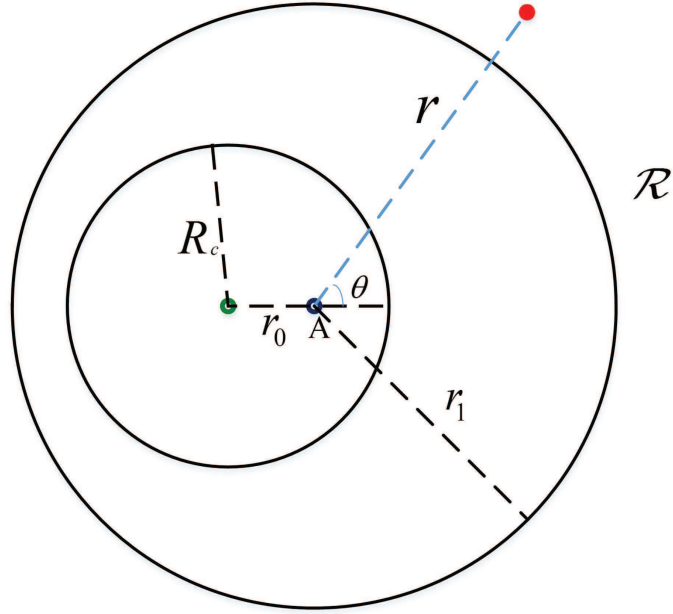}
\caption{Illustration of the system model when $r_1\geq R_c+r_0$}
\label{Proof_Lap_case2}
\end{figure}

Define
\begin{align}
 Q=\int_{\mathcal{R}(r_0,r_1)}\left(1-\frac{1}{1+\frac{s}{||y-x_0||^{\alpha_N}}}\right)\,dy,
\end{align}
then the remaining task is to evaluate $Q$.
By treating $x_0$ as the origin and changing to polar coordinates, $Q$ can be expressed as follows:
\begin{align}
Q=\iint_{\hat{\mathcal{R}}(r_0,r_1)}\left(1-\frac{1}{1+\frac{s}{r^{\alpha_N}}}\right)r\,drd\theta,
\end{align}
where $\hat{\mathcal{R}}(r_0,r_1)$ can be easily derived from ${\mathcal{R}}(r_0,r_1)$ and can be expressed
as:
\begin{align}
 \hat{\mathcal{R}}(r_0,r_1)=\left\{ (r,\theta) \big| r>r_1, r>z(\theta) \right\},
\end{align}
where $z(\theta)$ is the length of $\overline{AB}$ as shown in Fig. \ref{Proof_Lap_case1}, which can be easily
obtained by the law of cosine. It is worth pointing
out that the counterpart of the constraint $r>z(\theta)$ in (22) is $||y||>R_c$ in (19).
According to the relevant relationship of $r_0$ and $r_1$, the calculation of $Q$ can be divided
into the following two cases.
\begin{enumerate}
  \item Case I: $R_c-r_0<r_1<R_c+r_0$. In this case, the integration region
        $\hat{\mathcal{R}}(r_0,r_1)$ can
        be divided into two parts $\mathcal{R}_1$ and $\mathcal{R}_2$, i.e.,
        $\hat{\mathcal{R}}(r_0,r_1)=\mathcal{R}_1 \cup \mathcal{R}_2$, as shown in
        Fig. \ref{Proof_Lap_case1}. Mathematically, $\mathcal{R}_1=\left\{ (r,\theta)\big| r>r_1, |\theta|<\pi-\Theta\right\}$,
        and  $\mathcal{R}_2=\left\{ (r,\theta)\big| r>z(\theta), \pi-\Theta<|\theta|<\pi \right\}$.
        Then $Q$ can be evaluated as follows:
        \begin{align}\label{AE1}
         Q&=2\int_{0}^{\pi-\Theta}\int_{r_1}^{\infty}\left(
                  1-\frac{1}{1+\frac{s}{r^{\alpha_N}}}\right)r\,drd\theta
           +2\int_{\pi-\Theta}^{\pi}\int_{z(\theta)}^{\infty}\left(
                  1-\frac{1}{1+\frac{s}{r^{\alpha_N}}}\right)r\,drd\theta\\\notag
          &\overset{(a)}{=}2(\pi-\Theta)\frac{s^{\frac{2}{\alpha_N}}}{\alpha_N}
          \int_{\frac{1}{1+\frac{s}{r_1^{\alpha_N}}}}^{1}
              t^{\frac{2}{\alpha_N}-1}(1-t)^{-\frac{2}{\alpha_N}}\,dt\\\notag
              &\quad+2\int_{\pi-\Theta}^{\pi}\frac{s^{\frac{2}{\alpha_N}}}{\alpha_N}
          \int_{\frac{1}{1+\frac{s}{z^{\alpha_N}(\theta)}}}^{1}
              t^{\frac{2}{\alpha_N}-1}(1-t)^{-\frac{2}{\alpha_N}}\,dtd\theta\\\notag
          &\overset{(b)}{=}\frac{2(\pi-\Theta)s^{\frac{2}{\alpha_N}}}{\alpha_N}
           \bar{B}\left(\frac{1}{1+sr_1^{-\alpha_N}};\frac{2}{\alpha_N},1-\frac{2}{\alpha_N}\right)
           \\\notag
           &\quad+\frac{\Theta\pi s^{\frac{2}{\alpha_N}}}{N\alpha_N}\sum_{n=1}^{N}
          \sqrt{1-\theta_n^2}\bar{B}\left(\frac{1}{1+s(z(c_n))^{-\alpha_N}};\frac{2}{\alpha_N},1-\frac{2}{\alpha_N}\right),
        \end{align}
         where (a) follows from {\color{black}the step by using} $t=\frac{1}{1+\frac{s}{r^{\alpha_N}}}$, and (b) follows from
         {\color{black}the application of} Chebyshev-Gauss approximation.
  \item Case II: $r_1\geq R_c+r_0$. In this case, as shown in Fig. \ref{Proof_Lap_case2}, the integration region
                 $\hat{\mathcal{R}}(r_0,r_1)$ degrades to the following format
                 $\hat{\mathcal{R}}(r_0,r_1)=\left\{ (r,\theta)\big| r>r_1, |\theta|<\pi\right\}$.
                 Then $Q$ can be evaluated by following the similar steps as in Case I and the
                 following expression for $Q$ in Case II is obtained:
                 \begin{align}\label{AE2}
                  Q=\frac{2\pi s^{\frac{2}{\alpha_N}}}{\alpha_N}\bar{B}\left(\frac{1}{1+sr_1^{-\alpha_N}};\frac{2}{\alpha_N},1-\frac{2}{\alpha_N}\right).
                 \end{align}
\end{enumerate}
Now the proof for Lemma 2 is complete.
\subsection{Proof for Lemma 3}
From the proof for Lemma 2, we know that $\eta(s)$ can be expressed by the following integration:
\begin{align}
\eta(s)=-\lambda\iint_{\hat{\mathcal{R}}(r_0,r_1)}\left(1-\frac{1}{1+\frac{s}{r^{\alpha_N}}}\right)r\,drd\theta.
\end{align}
Then exchanging the order of the derivative and integration, the $t$-th derivative can be
expressed as follows:
\begin{align}
\eta^{(t)}(s)=t!(-1)^t\lambda\iint_{\hat{\mathcal{R}}(r_0,r_1)}
              \frac{r^{\alpha_N+1}}{(u+r^{\alpha_N})^{t+1}}\,drd\theta.
\end{align}
By dividing the calculation into the two cases as in the proof for Lemma 2 and following the similar steps as in (\ref{AE1}) and (\ref{AE2}), the expressions in (\ref{Lap_d1}) and (\ref{Lap_d2})
are obtained and the proof for Lemma 3 is complete.
\section{}
\subsection{Proof for Proposition 1}
$C_{\mathcal{A}_1}$ can be evaluated as follows:
\begin{align}
 C_{\mathcal{A}_1}&\overset{(a)}{=}\mathbb{E}_{\Phi}\left\{\int_{x\in \mathcal{D}}
                          \mathbb{E}_s\left\{\mathbf{1}\left(x\in\mathcal{A}_1 | \Phi,x,s\right)\right\}\,dx\right\}\\\notag
                  &=\mathbb{E}_{\Phi}\left\{\int_{x\in \mathcal{D}}
                          \mathbb{E}_s\left\{\mathbf{1}\left(r_1<A_s(||x||)| \Phi,x,s\right)\right\}\,dx\right\}\\\notag
                  &\overset{(b)}{=}\int_{0}^{2\pi}\int_0^{R_c}\mathbb{E}_{\Phi,s}\left\{
                     \mathbf{1}(r_1<A_s(r_0)|\Phi,r_0,s)\right\}r_0\,dr_0d\theta\\\notag
                  &\overset{(c)}{=}\int_{0}^{2\pi}\int_0^{R_c}\mathbb{E}_s\left\{\text{Pr}\left(
                     r_1<A_s(r_0)|r_0,s\right)\right\}r_0\,dr_0d\theta\\\notag
                  &\overset{(d)}{=}2\pi\int_0^{R_c}
                    \sum_{s\in\{L,N\}}P_s(r_0)F_{r_0|r_1}(A_s(r_0))r_0\,dr_0,
\end{align}
where in (a), $s\in\{L,N\}$ is a random variable which
{\color{black}indicate whether} the link from the UE to the UAV is an LoS or an NLoS link, and the probability
of s is defined in (\ref{P_lOS}),
(b) follows {\color{black}by} changing the order of {\color{black}the} expectation and integration and  then changing
to polar coordinates,
(c) and (d) {\color{black}follow by} taking expectation with respect to $\Phi$ and $s$ in sequence.

By following the same method, the expression for $C_{\mathcal{A}_2}$ and $C_{\mathcal{A}_3}$
can be obtained and the proof is complete.
\subsection{Proof for Corollary 1}
Here, only the proof for the conclusion that $\bar{C}_{\mathcal{A}_1}$ increases with $\delta$
is provided. The other two conclusions can be proved by {\color{black}following similar steps}.

To prove $\bar{C}_{\mathcal{A}_1}$ {\color{black}increasing} with $\delta$ is equivalent {\color{black}to prove that}
$C_{\mathcal{A}_1}$ increases with $\delta$.
Note that, when $A_s(r_0) \leq R_c-r_0$, $F_{r_1|r_0}(A_s(r_0))=0$, which has no contribution
to $C_{\mathcal{A}_1}$. Thus it is necessary to rewrite the integration constraint in (\ref{C_A1}) as
follows:
\begin{align}\label{C_A12}
C_{\mathcal{A}_1}=2\pi\sum_{s\in\{L,N\}}\int_{r_s(\delta)}^{R_c}
                    P_s(r_0)F_{r_0|r_1}(A_s(r_0))r_0\,dr_0,
\end{align}
where $r_s(\delta)=0$ when $\delta^{\frac{1}{\alpha_N}}H^{\frac{\alpha_s}{\alpha_N}}>R_c$,
otherwise $r_s(\delta)$ is the root of the equation:
$
A_s(r_0)=R_c-r_0
$.
{\color{black}Note that} it is easy to prove that the root always exists in $[0,R_c)$.

Now in (\ref{C_A12}), the integral function $F_{r_1|r_0}(A_s(r_0))$ is always {\color{black}positive}
and hence is an increasing function with $A_s(r_0)$.  In this cases, it can be concluded that
$F_{r_1|r_0}(A_s(r_0))$ increases with $\delta$, since $A_s(r_0)$ increasing with
$\delta$.
Further by noting that,  for any $0<\delta_1<\delta_2<\delta$, $r_s(\delta_1)\geq r_s(\delta_2)$,
the proof for  $\bar{C}_{\mathcal{A}_1}$ {\color{black}increasing} with $\delta$ is complete.
\subsection{Proof for Corollary 2}
We only prove that $\bar{C}_{\mathcal{A}_1}$ increases with $\lambda$
{\color{black}where the case that}
$\bar{C}_{\mathcal{A}_3}$ decreases with $\lambda$ can be proved similarly.

From (\ref{r1_CDF}), it is obvious that $F_{r_1|r_0}(r)$ increases with $\lambda$ when
$F_{r_1|r_0}(r)>0$.  Besides, from the last subsection, $C_{\mathcal{A}_1}$ can be expressed
as shown in (\ref{C_A12}), where $F_{r_1|r_0}(A_s(r_0))$ is always {\color{black}positive} in the
integration region. Based on the above two observations, it is proved that $C_{\mathcal{A}_1}$
increases with $\lambda$ and hence  $\bar{C}_{\mathcal{A}_1}$ increases with $\lambda$.

\section{Proof for Lemmas $4$-$6$}
\begin{enumerate}
  \item When $\text{UE}\in \mathcal{A}_1$, the conditional coverage probability can be calculated as follows:
\begin{align}
P^{1}(r_0,r_1)&=\text{Pr}\left(\text{SIR}_1>\epsilon\right)\\\notag
              &=\text{Pr}\left(|g_1|^2 > r_1^{\alpha_N}\epsilon(h_0+I_2)\right)\\\notag
              &\overset{(a)}{=}\mathbb{E}\left\{\text{exp}(-r_1^{\alpha_N}\epsilon(h_0+I_2))\right\}\\\notag
              &\overset{(b)}{=}\mathbb{E}_{g_0}\left\{\text{exp}\left(-\frac{r_1^{\alpha_N}\epsilon|g_0|^2}{(H^2+r_0^2)^{\alpha_L/2}}\right)\right\}
                 \mathbb{E}_{I_2}\left\{\text{exp}(-r_1^{\alpha_N}\epsilon I_2)\right\},
\end{align}
where (a) follows from {\color{black}the fact that} $g_1$ is Rayleigh distributed and (b) follows from {\color{black}the fact that} $h_0$ and
$I_2$ are independent random variables.  Finally, note that $|g_0|^2$ is a normalized
Gamma distribution with parameter $m_L$. {\color{black}Therefor, by applying} the Laplace transform of $I_2$ given in Lemma 1,
Lemma 4 is proved.
  \item When $\text{UE}\in \mathcal{A}_2$, the conditional coverage probability can be
      expressed as follows:
        \begin{align}
           P^{2}(r_0,r_1)&=\text{Pr}\left(\text{SIR}_2>\epsilon\right)\\\notag
              &=\text{Pr}\left(h_0+h_1> \epsilon I_2\right).
        \end{align}

        To calculate $P^{2}(r_0,r_1)$, we need to obtain the CDF for $h\overset{\Delta}{=}h_0+h_1$.
        Note that, $|g_0|^2 \sim \text{Gamma}(m_L,m_L)$, it is easily obtained that
        \begin{align}
           h_0 \sim \text{Gamma}(m_L,m_L(H^2+r_0^2)^{{\alpha_L}/2})\overset{\Delta}{=}
          \text{Gmama}(\alpha_0,\beta_0).
        \end{align}
        Similarly, we have
        \begin{align}
           h_1 \sim \text{Gamma}(1,r_1^{\alpha_N})\overset{\Delta}{=}
                     \text{Gamma}(\alpha_1,\beta_1).
        \end{align}
        Then the Laplace transform for $h$ can be expressed as follows:
        \begin{align}
         \mathcal{L}_h(s)&=\mathcal{L}_{h_0}(s)\mathcal{L}_{h_1}(s) \\\notag
                         &=\frac{\beta_0^2\beta_1^2}
                                {(s+\beta_0)^{\alpha_0}(s+\beta_1)^{\alpha_1}}\\\notag
                         &=\sum_{j=0}^1\sum_{k=1}^{\alpha_j} \frac{A_{jk}}{(s+\beta_j)^k},
        \end{align}
        where the last step follows from partial fraction decomposition. By taking the inverse
        Laplace transform, the CCDF for $h$ can be obtained as follows:
        \begin{align}
         \bar{F}_h(x)=\sum_{j=0}^1\sum_{k=1}^{\alpha_j}\frac{A_{jk}}{\beta_j^k}
                         \sum_{l=0}^{k-1}\frac{(\beta_jx)^l}{l!}e^{-\beta_jx}.
        \end{align}
        Now the coverage probability can be expressed as follows:
        \begin{align}
         P^2(r_0,r_1)=\mathbb{E}_{I_2}\left\{
                                \sum_{j=0}^1\sum_{k=1}^{\alpha_j}\frac{A_{jk}}{\beta_j^k}
                                    \sum_{l=0}^{k-1}\frac{(\beta_j\epsilon I_2)^l}{l!}e^{-\beta_j \epsilon I_2}
                                \right\}.
        \end{align}
        By further noting that $\mathbb{E}_{I_2}\{I_2^l e^{-uI_2}\}=(-1)^l\mathcal{L}^{(l)}_
        {I_2|r_0,r_1}(u)$, Lemma 5
        is proved.
  \item When $\text{UE}\in \mathcal{A}_3$, the conditional coverage probability can be calculated as follows:
        \begin{align}
        P^{3}(r_0,r_1)&=\text{Pr}\left(\text{SIR}_3>\epsilon\right)\\\notag
         &=\text{Pr}\left(|g_0|^2 > (H^2+r_0^2)^{\alpha_L/2}\epsilon(h_1+I_2)\right)\\\notag
         &=\mathbb{E}\left\{
                      \sum_{l=0}^{m_L-1}\frac{\left[u(h_1+I_2)\right]^l}
                      {l!}e^{-u(h+I_2)}
                      \right\},
        \end{align}
        where $u=m_L(H^2+r_0^2)^{\alpha_L/2}\epsilon$ and the last step follows from that
        $|g_0|^2 \sim \text{Gamma}(m_L,m_L)$. By further noting that
        $\mathbb{E}_{g_1,I_2}\{(h_1+I_2)^l e^{-u(h_1+I_2)}\}=(-1)^l\mathcal{L}^{(l)}_{h_1+I_2|r_0,r_1}(u)$,
        Lemma 6 is proved.
\end{enumerate}
\bibliographystyle{IEEEtran}
\bibliography{IEEEabrv,ref}
\end{document}